\documentclass[twocolumn,tighten]{aastex61}

\usepackage[]{natbib,epsfig,graphicx,rotating,url,bold-extra,xcolor}

\def\micron{$\mu$m}

\def\ltsim{\raisebox{-.4ex}{$\stackrel{<}{\sim}$}}
\def\gtsim{\raisebox{-.4ex}{$\stackrel{>}{\sim}$}}
\def\arcsec{$^{\prime\prime}$}

\newcommand{\skipthis}[1]{}

\newcommand{\mdotenv}{\mbox{$\dot{M}_{\rm env}$}}

\newcommand{\vff}{v_{f \mkern-3mu f}}

\shorttitle{Internal Luminosities of Low Mass Protostars}
\shortauthors{Huard \& Terebey}

\begin{document}

\title{The Power of SOFIA/FORCAST in Estimating \\
Internal Luminosities of Low Mass Class 0/I Protostars}

\author{Tracy L. Huard}
\affiliation{Department of Astronomy, University of Maryland, College Park, MD 20742, USA}
\affiliation{Columbia Astrophysics Laboratory, Columbia University, New York, NY 10027, USA}
\author{Susan Terebey}
\affiliation{Department of Physics and Astronomy, California State University at Los Angeles, Los Angeles, CA 90032, USA}

\begin{abstract}
With the Stratospheric Observatory for Infrared Astronomy (SOFIA) routinely operating science flights, we demonstrate that observations with the Faint Object infraRed CAmera for the SOFIA Telescope (FORCAST) can provide reliable estimates of the internal luminosities, $L_{\rm int}$, of protostars. We have developed a technique to estimate $L_{\rm int}$ using a pair of FORCAST filters: one ``short-wavelength'' filter centered within 19.7--25.3~$\mu$m, and one ``long-wavelength'' filter within 31.5--37.1~$\mu$m. These $L_{\rm int}$ estimates are reliable to within 30--40\% for 67\% of protostars and to within a factor of 2.3--2.6 for 99\% of protostars.  The filter pair comprised of F\,25.3\,$\mu$m and F\,37.1\,$\mu$m achieves the best sensitivity and most constrained results. We evaluate several assumptions that could lead to systematic uncertainties. The OH5 dust opacity matches observational constraints for protostellar environments best, though not perfectly; we find that any improved dust model will have a  small impact of 5--10\% on the $L_{\rm int}$ estimates. For protostellar envelopes, the TSC84 model yields masses that are twice those of the Ulrich model, but we conclude this mass difference does not significantly impact results at the mid-infrared wavelengths probed by FORCAST. Thus, FORCAST is a powerful instrument for luminosity studies targeting newly discovered protostars or suspected protostars lacking detections longward of 24~$\mu$m. Furthermore, with its dynamic range and greater angular resolution, FORCAST may be used to characterize protostars that were either saturated or merged with other sources in previous surveys using the \emph{Spitzer Space Telescope} or \emph{Herschel Space Observatory}.
\end{abstract}


\section{Introduction}

The \emph{Spitzer Space Telescope} enabled large infrared surveys of nearby star-forming molecular clouds yielding a census of young stellar objects (YSOs) in each cloud.  In particular, two \emph{Spitzer} legacy projects, ``From Molecular Cores to Planet-Forming Disks'' (c2d; \citealt{evans2003}) and ``Gould's Belt'' (GB), observed star-forming regions in 18 molecular clouds, resulting in the identification of 2966 YSO candidates, including 326 protostellar (Class 0/I) candidates (\citealt{dunham2015}).  Two other \emph{Spitzer} legacy projects were focused on the large star-forming regions of the Taurus (\citealt{rebull2010}) and Orion (\citealt{megeath2012}) molecular clouds, within which more than 3800 YSO candidates, including at least 500 protostellar candidates, were identified.

Since these \emph{Spitzer} surveys, some studies using the \emph{Herschel Space Observatory} --- including the ``Herschel Gould Belt Survey'' (\citealt{andre2010}) --- have been published identifying more protostars (e.g., \citealt{sadavoy2014}; \citealt{stutz2013}; \citealt{harvey2013}; \citealt{maury2011}).  These additional protostars generally represent a small ($\ltsim$5--10\%; e.g., \citealt{dunham2014}) increase in the number of Class 0/I protostars identified with \emph{Spitzer}, but they include ``extreme Class 0'' protostars, likely representing an earlier formation stage (\citealt{dunham2014}; \citealt{stutz2013}).

Among the most straightforward observational characteristics of protostars to derive is the bolometric luminosity, provided the spectral energy distributions (SEDs) are sufficiently covered, especially in the far-infrared and submillimeter regimes that dominate the emission.  However, many protostars have not been observed at these wavelengths and, if they have, the observations may lack the angular resolution necessary to reliably characterize the thermal emission from dust in the protostellar envelope.  Furthermore, the bolometric luminosity is ``contaminated'' by external heating by the interstellar radiation field; the {\it{internal}} (photospheric and accretion) luminosity, $L_{\rm int}$, better represents an intrinsic property of the protostar. Differences between bolometric and intrinsic luminosities tend not to be significant for typical or high luminosity protostars; those with luminosities $\ltsim$1.0~$L_\odot$ are most affected by external heating (e.g., \citealt{whitney2013}; \citealt{dunham2008}; \citealt{evans2001}).  \cite{dunham2008}  found that fluxes at 70~\micron\ alone were reliable indicators of $L_{\rm int}$.

\emph{Spitzer} and \emph{Herschel} surveys provided 70~$\mu$m fluxes for protostars, which may be used to estimate their internal luminosities.  However, many protostars either lack 70~\micron\ observations, or these observations suffer from insufficient dynamic range or angular resolution. With \emph{Spitzer} and \emph{Herschel} no longer obtaining such observations, a different approach is necessary to derive these estimates.  We therefore use radiative transfer models to investigate, in a manner similar to that of \cite{dunham2008}, the relationships between internal luminosities and FORCAST mid-infrared fluxes, which provide better dynamic range and angular resolution.  We demonstrate that FORCAST observations are sufficient  to estimate internal luminosities of protostars with reliability comparable to that achieved by 70~\micron\ observations.  In \S\ref{models}, we summarize the protostar models used in this study.  We discuss in \S\ref{forcast} the relevant characteristics of FORCAST imaging observations adopted to survey these models.  We present in \S\ref{results} results from these models, which confirm consistency with previous studies; we characterize relationships between observed FORCAST fluxes and internal luminosities of protostars.  In \S\ref{discussion}, we discuss the applicability and limitations of our results, and how these results may be used to further investigate low-mass protostars in nearby star-forming environments.  We summarize our findings in \S\ref{summary}.

\section{RADIATIVE TRANSFER MODELS}\label{models}

We employed the three-dimensional radiative transfer code, \textsc{Hochunk3d}, for protostars, developed by \cite{whitney2013} based on the two-dimensional version (\citealt{whitney2004b}; \citealt{whitney2003b}; \citealt{whitney2003a}) widely used in previous infrared surveys of protostars (e.g., \citealt{stutz2013}; \citealt{samal2012}; \citealt{carlson2011}; \citealt{forbrich2010}; \citealt{gramajo2010}; \citealt{enoch2009}; \citealt{merin2008}; \citealt{whitney2008}; \citealt{poulton2008}; \citealt{seale2008}; \citealt{simon2007}; \citealt{chapman2007}; \citealt{carlson2007}; \citealt{harvey2007}; \citealt{hatchell2007}; \citealt{tobin2007}; \citealt{bolatto2007}; \citealt{haisch2006}; \citealt{Kyoung2005}).  While \textsc{Hochunk3d} is equipped to deal with spiral and warp structures, and gaps in the disk, our current study is focused on the two-dimensional structures of protostellar disks and envelopes.  

Following \cite{dunham2008}, who used the \textsc{RadMC} code (\citealt{dullemond2004}) to model protostars observed with {\it{Spitzer}} IRAC (3--8~\micron; \citealt{fazio2004}) and MIPS (24, 70~\micron; \citealt{rieke2004}), we considered 350 models of typical protostars and flared disks within rotationally flattened protostellar envelopes, heated by external interstellar radiation fields (ISRFs), with assumed properties as summarized in this section.  For each model, we obtained results for 10 inclinations, $i$, uniformly spaced between $\cos{i}$ of 0 (edge-on disk) and 1 (face-on disk), or $\cos{i} = [0.05, 0.15, 0.25, ..., 0.95]$; thus, 3500 SEDs were constructed with a distribution of inclinations reflecting that expected for real protostars randomly oriented.  To limit statistical variations in the emergent fluxes, each model followed 10, 40, or 160 million photons, whichever was sufficient to yield signal-to-noise ratios (SNRs) of at least 5 at all inclinations and wavebands considered in this study, where SNRs were computed by \textsc{Hochunk3d} following \cite{wood1996}.  

The protostars emit as blackbodies at temperature 3000~K with randomly selected (uniformly, in log space) luminosities in the range 0.03--30 L$_\odot$, extending to more luminous protostars than \cite{dunham2008}.  As mentioned in \cite{crapsi2008}, the precise temperature assumed for the protostars is not critical since all of the emission is reprocessed by the disks and envelopes.

The flared protostellar disks have a density structure, $\rho_{\rm disk}$, that decreases as a power law in the midplane radially ($\varpi$) while decreasing exponentially perpendicular to the midplane ($z$) according to (e.g., \citealt{shakura1973}; \citealt{lazareff1990}; \citealt{pringle1981}; \citealt{bjorkman1997}; \citealt{hartmann1998}; \citealt{whitney2003a}):
\begin{equation}
\rho_{\rm disk} = \rho_{d0} \left(1\hspace{-2pt}-\hspace{-2pt}\sqrt{\frac{R_*}{\varpi}}\right)\left(\frac{R_*}{\varpi}\right)^{\beta+1}\hspace{-6pt}\exp{\left \{ -\frac{1}{2} \left[ \frac{z}{h(\varpi)} \right]^2 \right \}} ,
\end{equation}

\noindent
where $R_*$ is radius of the protostar, and the scale height increases as a power law, $h(\varpi) \propto ( \varpi / R_* )^{\beta}$ with $\beta = 9/7$, which is consistent with a self-irradiated passive disk (\citealt{chiang1997}).  No accretion energy is considered in the disks.  The disks have inner radii given by the sublimation temperature and outer radii of 100 AU, where the scale height is 20 AU.  The disk masses, which set the overall density normalizations, $\rho_{d0}$, are randomly selected (uniformly, in log space) in the range $10^{-5}-10^{-3}$ M$_\odot$.

The rotationally flattened envelopes have density profiles, $\rho_{\rm env}$, that may be parameterized in terms of the centrifugal radius, $R_c$, and the polar angle, $\theta_0$, of the streamline of infalling material at large radial distances, $r$ (\citealt{cassen1981}; \citealt{ulrich1976}):
\begin{eqnarray}
\rho_{\rm env} =  \rho_o \left( \frac{R_c}{r} \right)^{3/2}  
              \left( 1 + \frac{\mu}{\mu_0} \right)^{-1/2} \left( \frac{\mu}{ \mu_0} + \frac{2 \mu^2_0 R_c}{r} \right)^{-1} ,
\label{rhoenvLONGEQN} 
\end{eqnarray}

\noindent
where $\mu \equiv \cos{\theta}$ and $\mu_0 \equiv \cos{\theta_0}$. The constant $\rho_o$ is defined by
\begin{eqnarray}
\rho_{o} & = & \frac{\dot{M}_{\rm env}}{4 \pi \sqrt{G (M_* + M_{\rm disk}) R_c^3} } \\
& \approx & \frac{\dot{M}_{\rm env}}{4 \pi \sqrt{G M_* R_c^3}} ,
\end{eqnarray}

\noindent
where $\dot{M}_{\rm env}$ is the mass infall rate in the envelope, $M_*$ is the mass of the central protostar, $M_{\rm disk}$ is the mass of the disk, and $M_{\rm disk} \ll M_*$.  The Ulrich profile assumes the gas is in free-fall towards a fixed central mass.  While the Ulrich profile is typically adopted for the entire envelope, as we have also done in our study, we remind readers that it most accurately reflects free-fall envelope densities at radial distances, $r$, within which the  mass is dominated by the central protostar rather than the disk or envelope.  Thus, the Ulrich profile deviates from an accurate collapse profile as the envelope mass interior to $r$ increases appreciably relative to $M_*$ (e.g., \citealt{shu1977}), which is likely the case in real protostellar envelopes (see \S\ref{discussion}).

In this formulation, the three input parameters $\dot{M}_{\rm env}, M_*$, and  $R_c$ suffice to specify the envelope density profile. The parameters $\dot{M}_{\rm env}$ and $M_*$ are related to the density normalization and collapse timescale. The parameter $R_c$ is related to rotation and is often set equal to the disk radius. A fourth (less important) parameter arises because of the necessity to set a maximum cloud envelope radius, ${R}_{\rm env}$, in order to compute a model. Other formulations of the Ulrich profile are present in the literature; for example, \cite{furlan2016} preferred to express the profile in terms of the envelope density at a fiducial radial distance (1000~AU), assuming $M_* = 0.5~$M$_\odot$.  We instead have recast Equation~\ref{rhoenvLONGEQN} in terms of ${M}_{\rm env}$, which is often used in the literature, and use it to set the normalization instead of $\dot{M}_{\rm env}$, by noting that the streamlines become radial at large $r$, so that in the limit $r /R_c \to \infty $ and $\mu / \mu_0 \to 1$  the envelope density simplifies to become
\begin{eqnarray}
\rho_{\rm env} & \simeq & \rho_{o} \left( \frac{R_c}{r} \right)^{3/2}  2^{-1/2} .
\label{rhoenvEQN} 
\end{eqnarray}

\noindent
The mass of the envelope is then given by
\begin{eqnarray}
M_{\rm env} & = & 4 \pi \int \rho_{\rm env} ~r^2 dr   \\
\label{MenvEQNintegral} 
& = & \frac{2}{3}    \frac{\dot{M}_{\rm env} R_{\rm env}^{3/2}} {(2 G M_*)^{1/2} }.
\label{MenvEQN} 
\end{eqnarray}

\noindent
In order to facilitate comparison with the \cite{dunham2008} and \cite{crapsi2008} results, we adopt $M_* = 0.5~M_{\odot}$ for the model suite. Similarly, we consider envelopes with outer radii of 14,000 AU, envelope masses randomly selected (uniformly, in log space) in the range 1--10 M$_\odot$, $R_c$ randomly selected in the range 100--900~AU, and bipolar cavities (created from protostellar outflows) with shape following the streamline with opening angle of 15$^\circ$; the density within each cavity is set to the density of the outermost region of the envelope (\citealt{dunham2008}).

As discussed in \cite{whitney2013}, the external ISRF adopted by default in \textsc{Hochunk3d} is that found by \cite{mathis1983} for the solar neighborhood, while   \cite{dunham2008} adopted that of \cite{black1994} modified at ultraviolet wavelengths for consistency with \cite{draine1978}.  \cite{evans2001} discuss differences between these ISRFs, though we note that the default ISRF in \textsc{Hochunk3d} does not include the cosmic background component dominating at millimeter wavelengths.  For consistency with \cite{dunham2008}, we adapted \textsc{Hochunk3d} to use the ``Black-Draine'' ISRF in our current study.  To account for environmental differences among protostellar envelopes, including differing amounts of dust in the molecular clouds surrounding these envelopes, the strength of the ISRF was adjusted by a scale factor and then attenuated and reddened.  For each envelope, this scale factor is randomly selected in the range $1/3-3$, distributed logarithmically about unity, and the dust visual extinction is randomly selected in the range 1-5 magnitudes.

\subsection{Dust Grain Properties}\label{dustsection}

\begin{figure}[t!]
\begin{center}
\begin{turn}{90}
\includegraphics[trim=0in 0in 0.5in 0.5in, height=3.5in]{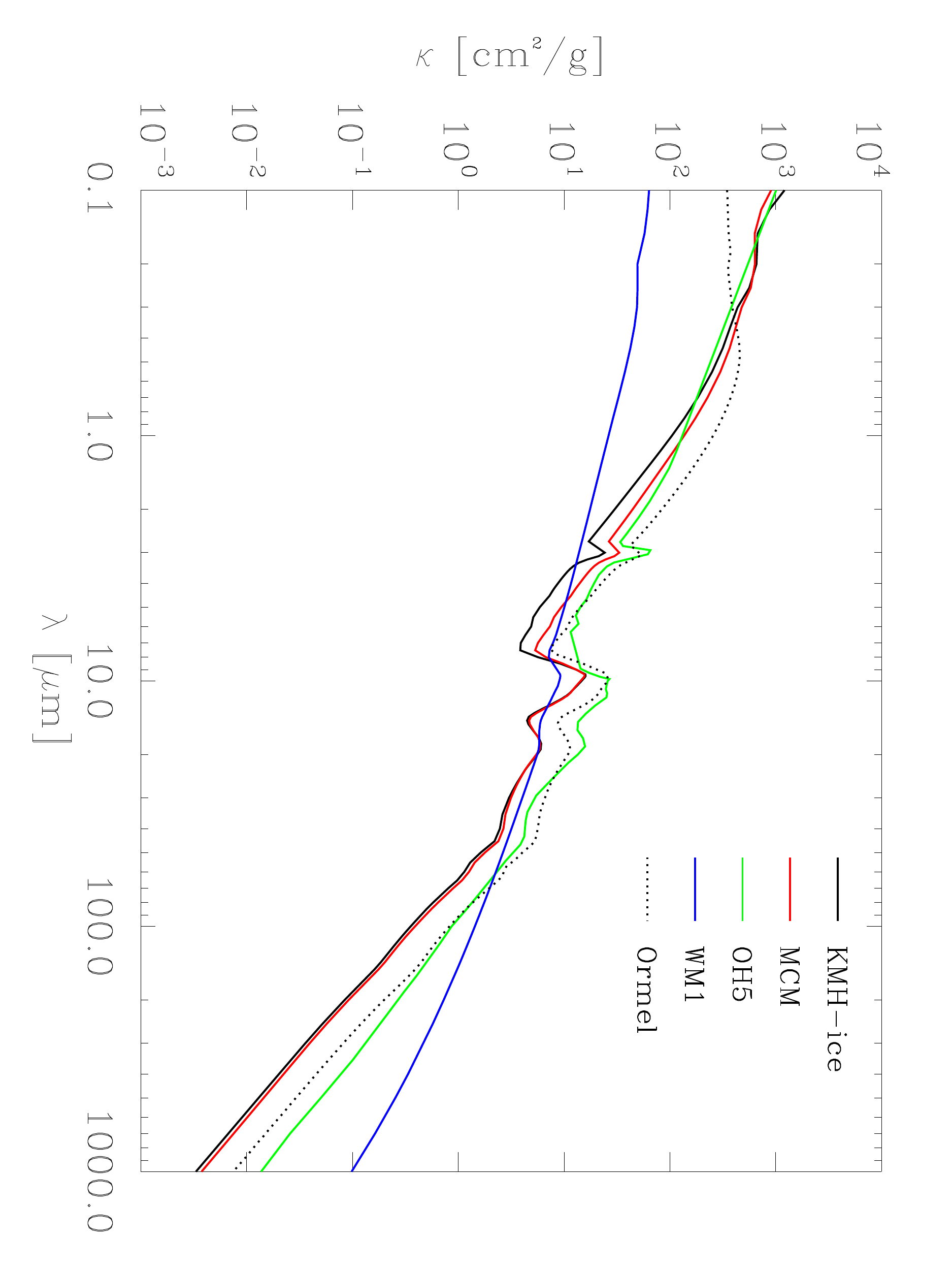}
\end{turn}
\begin{turn}{90}
\includegraphics[trim=0.5in 0in 0.5in 0.5in, height=3.5in]{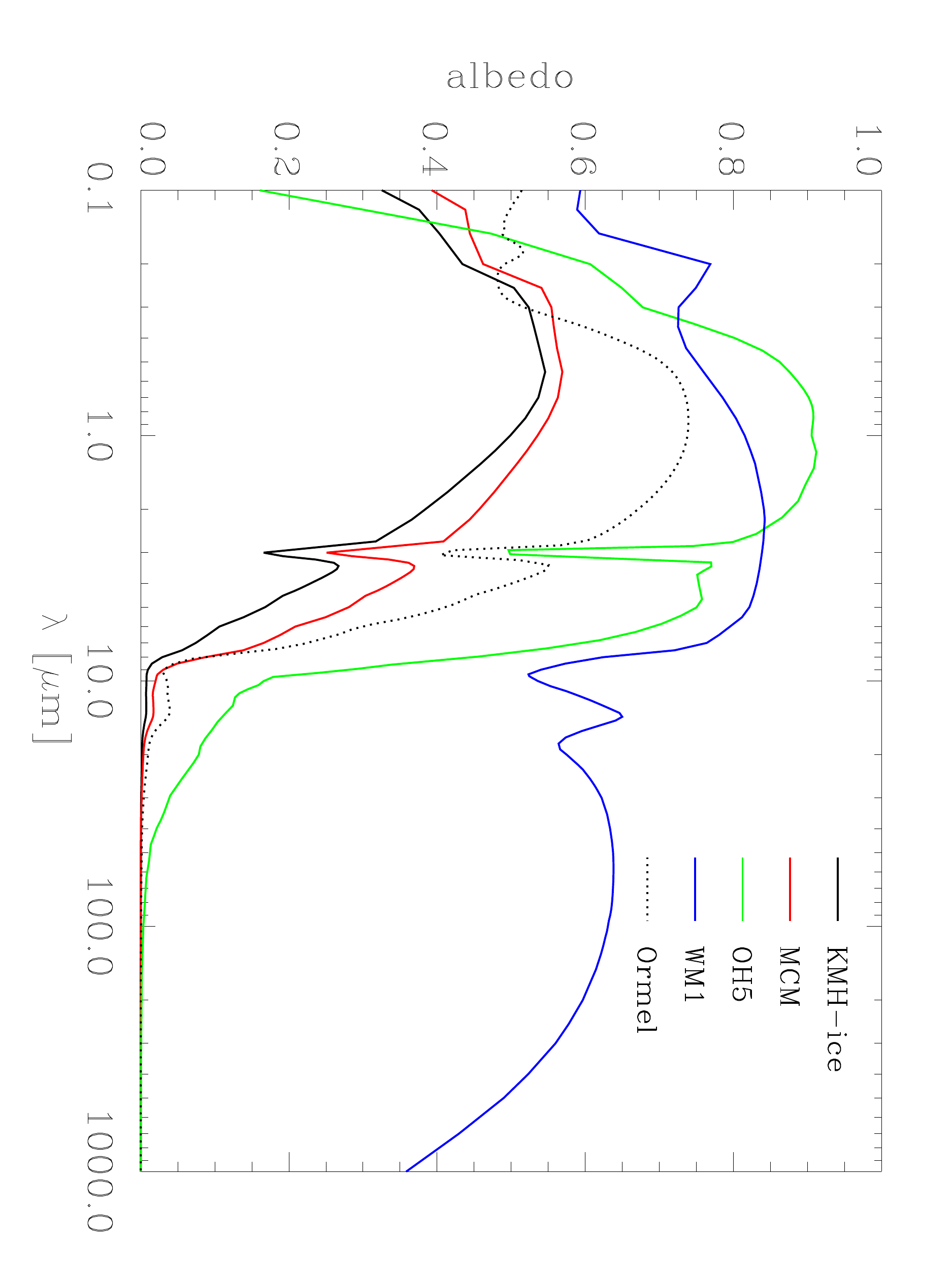}
\end{turn}
\end{center}
\caption{Opacities (top) and albedos (bottom) of the gas and dust mixture, assuming a gas-to-dust mass ratio of 100, for the different grain populations considered in this study.
\label{grainprops}} 
\end{figure}

The optical properties of the envelope dust adopted by \cite{dunham2008} were not available; therefore, we experimented with different dust grain populations available in the literature.  The first three grain populations that we considered were readily available in the \textsc{Hochunk3d} distribution.  The first population, which we refer to as ``KMH-ice'' dust, was that found by \cite{kmh1994} for the average Galactic interstellar medium, except with water-ice mantles making up the outer 5\% (in radius) of the grains.  The second population that we tried was the ``molecular cloud model'' (hereafter, referred to as ``MCM'') dust appropriate for protostellar envelopes and described in \cite{whitney2013}.  The third dust population was ``model 1'' dust, which we refer to as ``WM1'' dust, used by \cite{wood2002} to model the disk of the classical T Tauri star HH 30 IRS.  A fourth population was the thinly ice-mantled, coagulated dust of \cite{oh1994}, often referred to as ``OH5'' grains in the literature (e.g., \citealt{evans2001}; \citealt{shirley2005}), augmented by the opacities of \cite{pollack1994} at wavelengths shorter than 1.25~\micron, as described in \cite{dunham2010evolve}.  The last population was that of \cite{ormel2011} adopted by \cite{furlan2016}, which includes a mixture of ice-coated silicate and bare graphite grains of radii 0.1--3~$\mu$m.   The OH5 and Ormel populations were not available in \textsc{Hochunk3d}, but we included them for this study.  For reference, the opacities and albedos for the five considered grain populations are shown in Figure~\ref{grainprops}.

\begin{figure}[t!]
\begin{center}
\includegraphics[trim=0.2in 0in 0in 0in, width=3.35in]{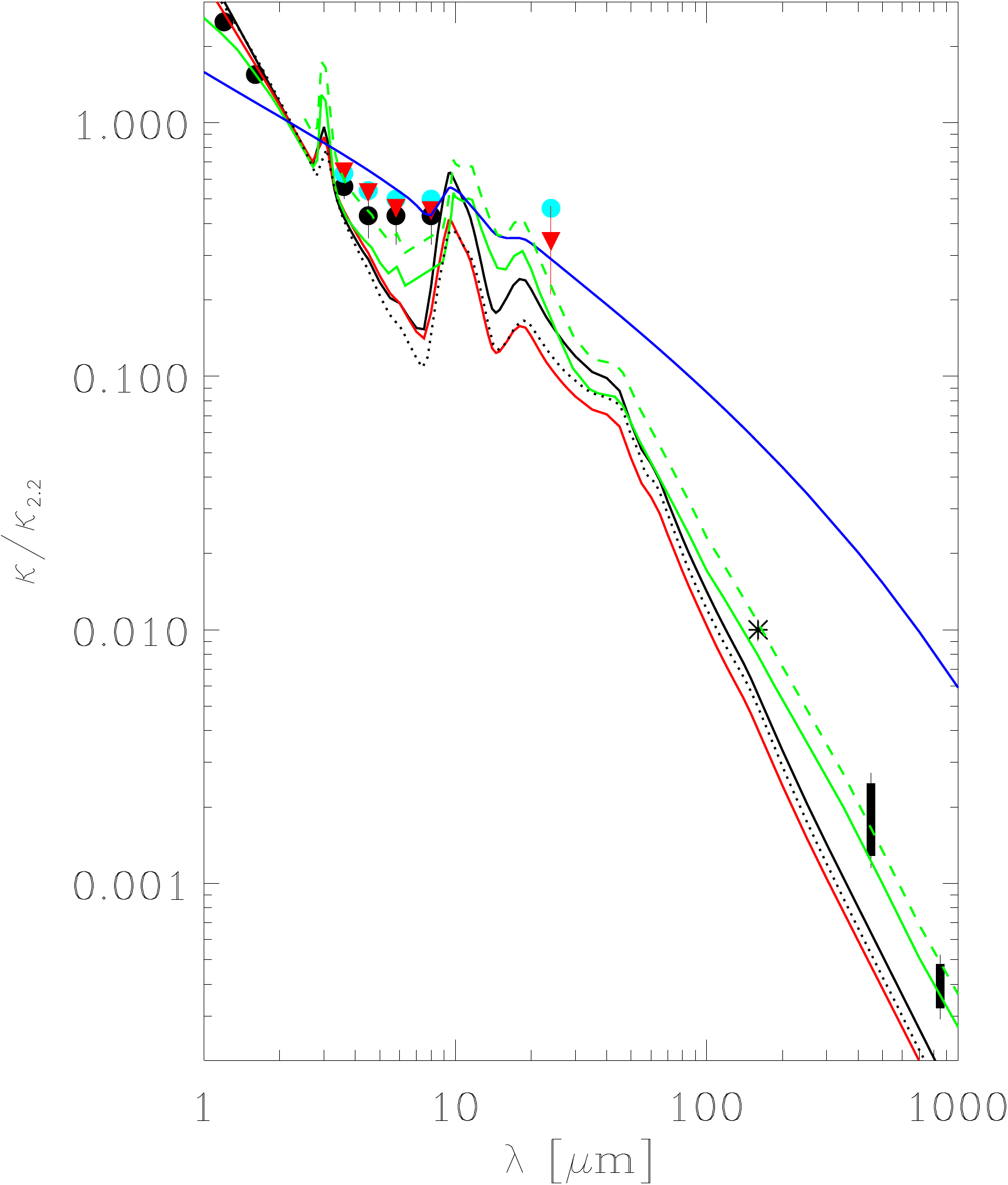}
\end{center}
\caption{Comparison of observationally derived dust opacities, relative to that at 2.2~$\mu$m, in protostellar environments and dense molecular clouds with those of grain populations considered in this study.  Observed relative opacities at 1.2--24~$\mu$m from \cite{indebetouw2005}, \cite{flaherty2007}, and \cite{chapmanDUST2009} are plotted as filled black circles, light blue circles, and red triangles; the 160~$\mu$m and 250~$\mu$m relative opacities from \cite{terebey2009} and \cite{suut2013} are plotted as an asterisk and open circle, respectively; and the relative opacity ranges at 450 and 850~$\mu$m from \cite{shirley2011} are plotted as vertical bars.  Error bars have been included, though in most cases they are covered by the symbol.  The relative opacities of grain populations are plotted as curves with the same color scheme as in Figure~\ref{grainprops}: KMH-ice (black), MCM (red), WM1 (blue), OH5 (green), Ormel (black dotted curve).  The dashed green line represents the relative opacity of the OH5 population, increased by 35\% for $\lambda \geq 2.5~\mu$m.
\label{obsgrainprops}} 
\end{figure}

Observationally derived, infrared and submillimeter dust opacities, for protostellar environments, relative to the opacity at 2.2~$\mu$m are shown in Figure~\ref{obsgrainprops}.  A comparison with the relative opacities from grain populations considered in this study suggests that the OH5 grains best reproduce these observations.  For this reason, we adopt the OH5 population in this study.

As evident in Figure~\ref{obsgrainprops}, none of the grain populations yield relative opacities at 1.2--850~$\mu$m that are fully consistent with observations.  Increasing the relative opacity of OH5 grains by 35\% for $\lambda \geq 2.5~\mu$m yields better agreement with observations. In order to obtain some handle on how a grain population better constructed for protostellar environments may affect our results, we rerun the models with these ``revised OH5'' opacities.  We stress, however, that artificially increasing the mid-infrared and submillimeter relative opacities of the OH5 grains is not consistent with element abundance constraints of grain populations; such opacities would result from larger grains, and inclusion of these larger grains would necessarily come at the expense of smaller grains to conserve element abundances.  Constructing a protostellar grain population is beyond the scope of this study.

\section{FORCAST Filters and Sensitivities}\label{forcast}

The FORCAST instrument (\citealt{herter2012}; \citealt{adams2010}) on SOFIA (\citealt{young2012}) obtains mid-infrared images and spectra at 5.4--37.1~\micron\ on two detectors: the short-wavelength channel (SWC) and the long-wavelength channel (LWC).  Using a dichroic, these channels simultaneously image two wavebands; alternatively, a single channel may be used to directly image one waveband.  For our study, we consider only FORCAST images using the seven filters, listed in Table~\ref{sensitivity}, in the range 19.7--37.1~\micron\ for typical observing conditions:  specifically, an altitude of 41,000 feet, 7.1~\micron\ of precipitable water vapor at the zenith, and telescope pointings at 50$^\circ$ from the zenith (e.g., \citealt{horn2001}).

\begin{table}[t!]
\caption{Fiducial Sensitivity Limits}
\label{sensitivity}
\begin{center}
\begin{tabular}{ccc}
\hline
\hline  \\[-9pt]
FORCAST	& Dichroic	& Direct	\\
Filter$^a$		& [mJy]	& [mJy]	\\
\hline \\[-9pt]
F\,19.7\,$\mu$m (SWC)	& 25		& 23\\		
F\,24.2\,$\mu$m (LWC)	& ...		& 50		\\
F\,25.3\,$\mu$m (SWC)	& 63		& 59		\\
F\,31.5\,$\mu$m (LWC)	& 84		& 60		\\
F\,33.6\,$\mu$m (LWC)	& 182	& 116	\\
F\,34.8\,$\mu$m (LWC)	& 114	& 78		\\
F\,37.1\,$\mu$m (LWC)	& 168	& 97		\\
\hline  \\[-9pt]
\end{tabular}
\end{center}
\footnotesize{{$^a$}For each filter, the channel is included in parentheses.}\\
\end{table}

For purposes of discussion, we adopt fiducial sensitivity limits as those point source flux densities associated with SNR=3 after an hour exposure time.  Most FORCAST surveys of star-forming regions are likely to require greater SNRs achieved in reasonable times; thus, we expect most studies will focus on sources brighter than given by these limits.  Using the online SOFIA Instrument Time Estimator\footnote{\url{https://dcs.sofia.usra.edu/proposalDevelopment/SITE/}}, we determined these fiducial sensitivity limits, in typical observing conditions, for the FORCAST filters operating in direct and dichroic modes, as listed in Table~\ref{sensitivity}.

\begin{figure}[b!]
\begin{center}
\begin{turn}{90}
\includegraphics[trim=0in 0in 0in 0.2in, width=2.75in]{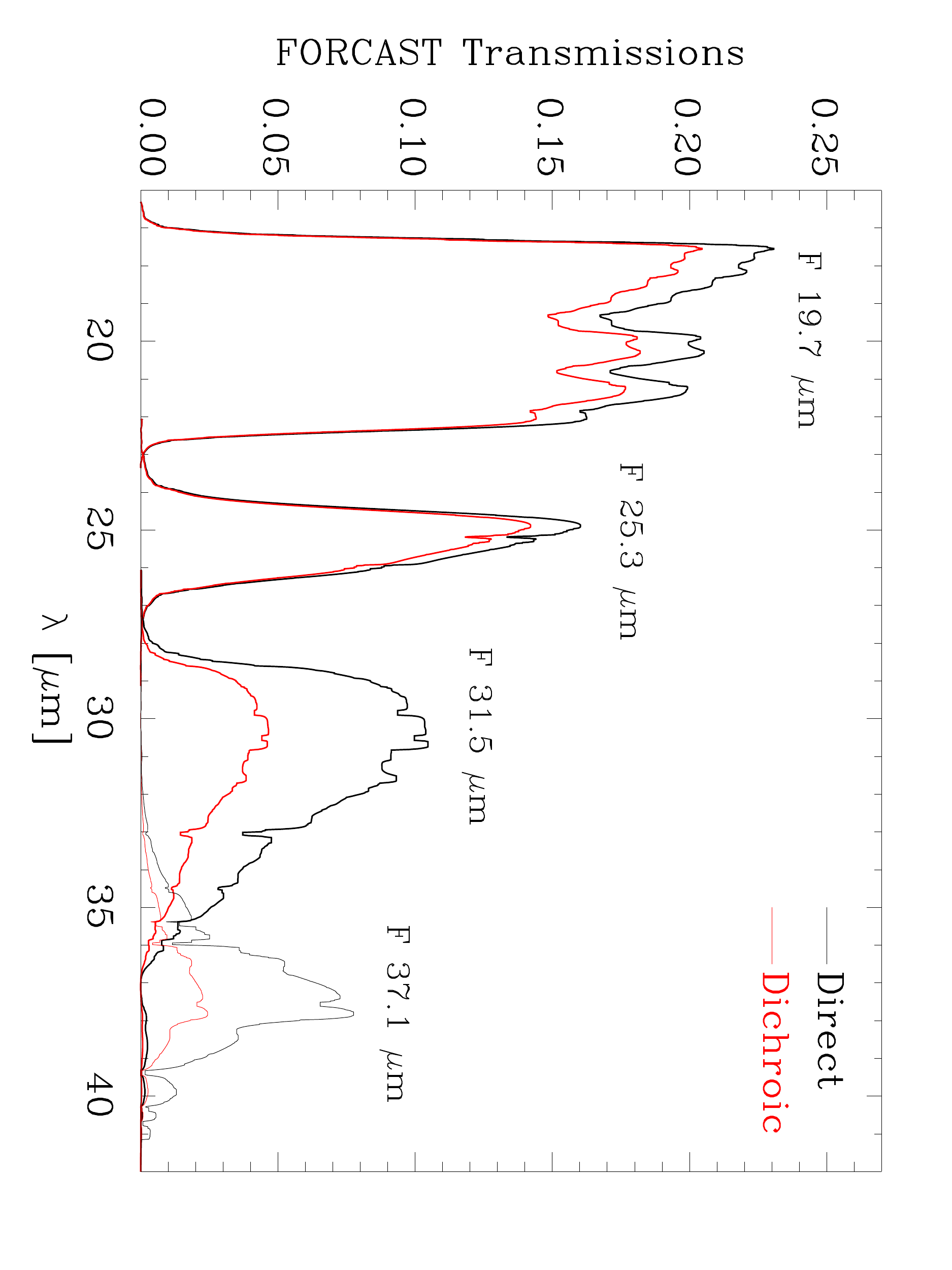}
\end{turn}
\end{center}
\caption{The effective transmission functions of four of the seven FORCAST filters considered in this study, assuming typical observing conditions.  The functions associated with observations obtained in direct and dichroic modes are plotted by the black and red curves, respectively, and account for absorption by the atmosphere and optical elements as well as the detector response.
\label{transmissions}} 
\end{figure}

Figure~\ref{transmissions} illustrates the effective transmissions of four of these FORCAST filters, accounting for the atmosphere, optics (e.g., the filter itself, optical blockers, and dichroic, as appropriate), and detector response.  Except for the F\,24.2\,$\mu$m filter, which operates only in direct mode, we included in \textsc{Hochunk3d} the dichroic transmission functions of the filters listed in Table~\ref{sensitivity} in order to derive FORCAST flux densities of protostellar models.  For the F\,24.2\,$\mu$m filter, we included the direct transmission function.  For each filter, the shapes of the direct and dichroic transmission functions are similar; the primary difference is in the overall scale factor of the transmission.  Therefore, no significant difference is expected in flux densities derived from dichroic and direct transmission functions, only in the observing time required to detect them, particularly for filters with effective wavelengths greater than 30~\micron.

\section{RESULTS}\label{results}

\begin{figure*}[t!]
\begin{center}
\includegraphics[trim=0.6in 1.5in 0.9in 0.2in, width=7.3in]{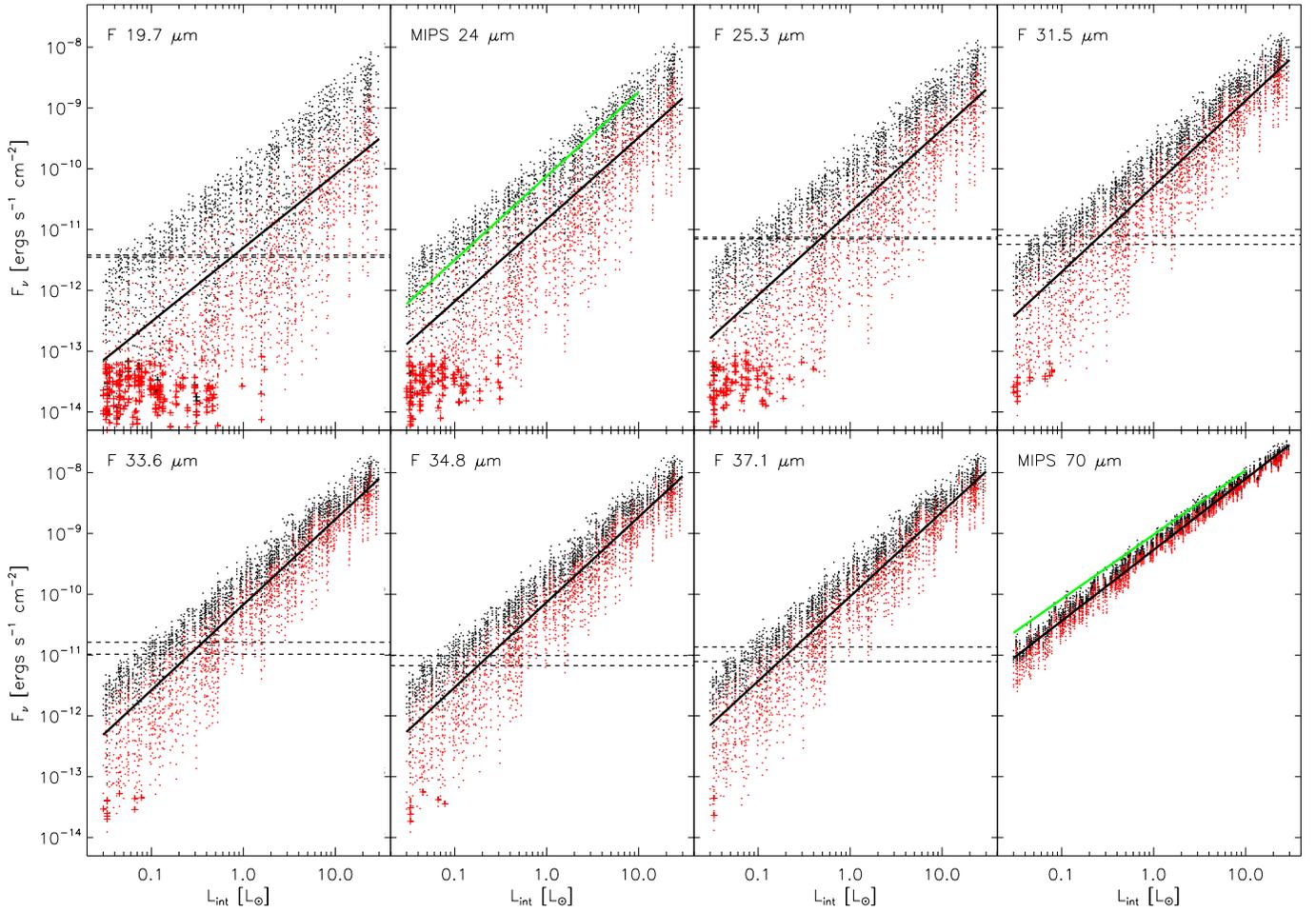}
\end{center}
\caption{MIPS and FORCAST fluxes as a function of $L_{\rm int}$ for protostellar disks and envelopes with OH5 dust.  Each FORCAST panel also includes horizontal dashed lines representing the fiducial sensitivity limits for that filter in dichroic (top dashed line) and direct (bottom dashed line) modes, as listed in Table~\ref{sensitivity}.  The black and red points represent our models with $\cos{i} \geq 0.5$ and $\cos{i} < 0.5$, respectively.  Those models dominated by photons originating from the ISRF (see \S\ref{discussion}) are identified by plus signs; FORCAST is not sensitive to such models.  Our best-fit lines to all models, excluding the ISRF-dominated models, are shown as black lines.  The MIPS 24~$\mu$m and 70~$\mu$m panels include the fits from \cite{dunham2008} as green lines, for reference.
\label{MIPSFORCASTpanels}} 
\end{figure*}

\begin{table*}[t!]
\caption{Single-Filter Fits}
\label{linearfits}
\begin{center}
\begin{tabular}{ccccc}
\hline
\hline  \\[-9pt]
Filter			& $m$				& $b$				& $\chi^2_{\rm red}$	& $\sigma_L^a$	\\
\hline \\[-9pt]
MIPS\,24\,\micron	& 1.35 $\pm$ 0.02		& $-$10.83 $\pm$ 0.01	& 298	& 0.56	\\
MIPS\,70\,\micron	& 1.169 $\pm$ 0.003	& $-$9.270 $\pm$ 0.002	& 10		& 0.12	\\
\hline \\[-9pt]
F\,19.7\,$\mu$m	& 1.21 $\pm$ 0.02		& $-$11.30 $\pm$ 0.02	& 477	& 0.55	\\
F\,24.2\,$\mu$m	& 1.36 $\pm$ 0.01		& $-$10.68 $\pm$ 0.01	& 256	& 0.39	\\
F\,25.3\,$\mu$m	& 1.36 $\pm$ 0.02		& $-$10.71 $\pm$ 0.01	& 278	& 0.38	\\
F\,31.5\,$\mu$m	& 1.41 $\pm$ 0.01		& $-$10.289 $\pm$ 0.009	& 143	& 0.29	\\
F\,33.6\,$\mu$m	& 1.40 $\pm$ 0.01		& $-$10.173 $\pm$ 0.008	& 121	& 0.26	\\
F\,34.8\,$\mu$m	& 1.400 $\pm$ 0.009	& $-$10.129 $\pm$ 0.008	& 112	& 0.25	\\
F\,37.1\,$\mu$m	& 1.391 $\pm$ 0.009	& $-$10.035 $\pm$ 0.007	& 97		& 0.24	\\
\hline  \\[-9pt]
\end{tabular}
\end{center}
\footnotesize{{$^a$$\sigma_L$ represents the dispersion between best-fit and input $\log{L_{\rm int}}$ values.  For MIPS filters, all models are considered; for FORCAST filters, only models detectable, given fiducial sensitivities in dichroic mode (except for F\,24.2\,$\mu$m, where we assumed direct mode), are considered.}}\\
\end{table*}

Following the approach of \cite{dunham2008} and adopting OH5 dust, as discussed in \S\ref{dustsection}, our results for FORCAST and MIPS fluxes, at a distance of 140\,pc, as a function of $L_{\rm int}$, are shown in Figure~\ref{MIPSFORCASTpanels}, demonstrating that $L_{\rm int}$ is best indicated by the 70~$\mu$m flux.  This figure illustrates increased scatter, particularly at smaller wavelengths and for lower luminosity protostars.  The increased scatter is primarily due to geometric effects of inclination.  Scatter introduced from inclination may be understood by referring to the SEDs of a standard Class~I protostar shown in Figure~14 of \cite{whitney2013}.  The flux at 70~$\mu$m is relatively unchanged with inclination, while fluxes at shorter wavelengths, particularly for $\lambda < 40~\mu$m, vary considerably for the same protostar observed at different inclinations.

Like \cite{dunham2008}, we derive fluxes  within 6\arcsec-radius (840~AU at 140~pc) apertures, which is the {\emph{Spitzer}} resolution (full width at half maximum; FWHM) at 24~$\mu$m, for $\lambda < 40~\mu$m and within 20\arcsec-radius (2800~AU at 140~pc) apertures at 70~$\mu$m. Typical resolutions achieved by FORCAST filters in 19.7--37.1~$\mu$m are 2.1--3.4\arcsec; thus, in principle, our aperture fluxes for FORCAST filters will capture a greater fraction of the total fluxes. At least for point sources, the 6\arcsec-radius aperture already captured most of the flux in Spitzer observations; any difference between 6\arcsec-radius aperture fluxes derived from Spitzer observations and those derived from FORCAST observations is expected to be negligible.

Using a linear least-squares fitting method, we determined the best-fit parameter values, $m$ and $b$, characterizing the dependence of flux, $F_\nu$ (measured in erg s$^{-1}$ cm$^{-2}$) at a distance 140\,pc, on $L_{\rm int}$ (measured in L$_\odot$):
\begin{equation}
\log{F_\nu} = m \log{L_{\rm int}} + b ,
\label{singlefit1}
\end{equation}

\noindent
where we note that observed photometry is typically given in terms of {\it{flux density}}, $S_\nu\,=F_\nu\,\nu^{-1}$.  The best fits are illustrated in Figure~\ref{MIPSFORCASTpanels}, and the associated parameter values are listed in Table~\ref{linearfits}, which also lists the standard reduced chi-squared $\chi_{\rm red}^2$ statistics for assessing the quality of these fits.  More directly meaningful is Column 5 of Table~\ref{linearfits}, which lists values for the dispersion ($\sigma$) between best-fit and input $\log{L_{\rm int}}$ values
\begin{equation}
\sigma_L \equiv\ \sigma \left[ \log{L_{\rm int}{\rm (fit)}} - \log{L_{\rm int}} \right],
\label{singlefit2}
\end{equation}

\noindent
where the best-fit values can be explicitly written, for clarity, as
\begin{equation}
\log{L_{\rm int}{\rm (fit)}} = \frac{\log{F_\nu} - b}{m} ,
\label{singlefit3}
\end{equation}

\noindent
using the values for $m$ and $b$ listed in Table~\ref{linearfits}.  These dispersions, $\sigma_{L}$, provide a direct means for quantifying the reliability of $L_{\rm int}$ estimates based on these fits.  For example, $\sigma_L = 0.12$ when using 70~$\mu$m fluxes; thus, $L_{\rm int}$ estimates based on these fluxes are reliable to within a factor of 1.3 for 67\% of the models (i.e., 1$\sigma$) and 2.3 for 99\% of the models (i.e., 3$\sigma$), assuming normal distributions.  In contrast, the 19.7~$\mu$m fluxes, which yield $\sigma_L = 0.55$, result in luminosities reliable only to within a factor of 45 (3$\sigma$; factor of 3.5 for 1$\sigma$).  Clearly, the capability that {\it{Spitzer}} and {\it{Herschel}} had in obtaining 70~$\mu$m fluxes was critical in characterizing protostars.

\begin{figure*}[t!]
\begin{center}
\includegraphics[trim=0in 0in 0in 0in, width=7.1in]{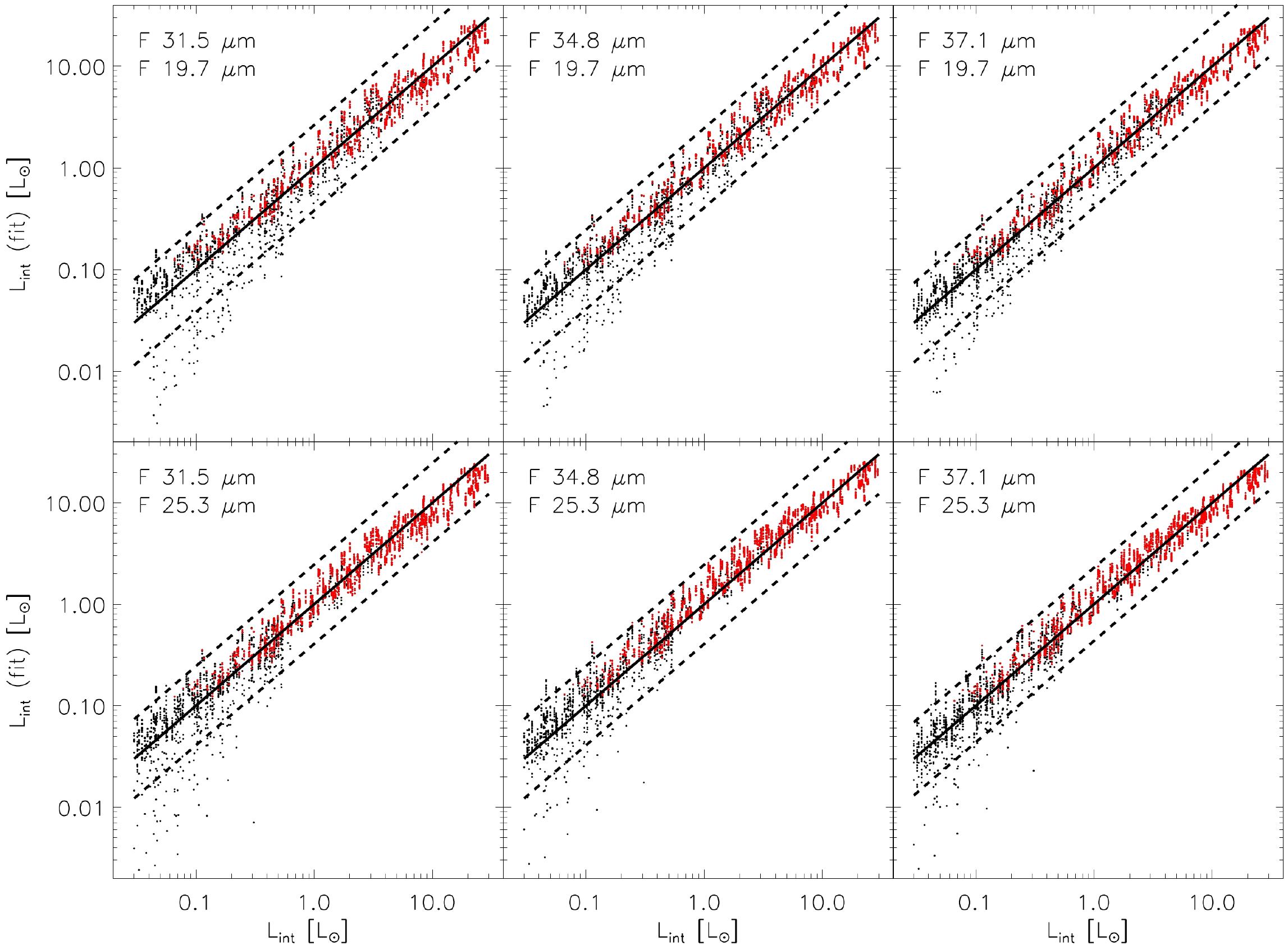}
\end{center}
\caption{Comparison of $L_{\rm int}$ estimates, derived from mid-infrared fluxes, with input $L_{\rm int}$ for half of the two-filter combinations considered in this study.  Plots associated with $L_{\rm int}$ estimates using long-wavelength filters F\,31.5\,$\mu$m, F\,34.8\,$\mu$m, and F\,37.1\,$\mu$m are plotted in the left, center, and right panels; those estimates using short-wavelength filters F\,19.7\,$\mu$m and F\,25.3\,$\mu$m are plotted in the top and bottom panels.  These plots are similar to those utilizing filters F\,24.2\,$\mu$m and F\,33.6\,$\mu$m, which are not included in this figure.  The black points represent models with a flux in at least one of the two relevant FORCAST bands less than the dichroic fiducial sensitivity limit; the red points represent models detectable within the fiducial 1-hour exposures.  ISRF-dominated models are not included.  The solid lines represent perfect agreement between the estimates and model input values, while the dashed lines represent estimates within 3$\sigma_L$ ranges.
\label{LintEstimates}} 
\end{figure*}

\begin{table*}[t!]
\caption{Two-filter fits}
\label{regressionfits}
\begin{center}
\begin{tabular}{llcccc}
\hline
\hline  \\[-9pt]
Filter 1			& Filter 2			& $C_1$				& $C_2$				& $C_3$	& $\sigma_L^a$ 	\\
\hline \\[-9pt]
F\,37.1\,$\mu$m	& F\,19.7\,$\mu$m	& 1.032 $\pm$\ 0.005	& $-$0.322 $\pm$\ 0.004	& 6.671	& 0.13	\\ 
F\,37.1\,$\mu$m	& F\,24.2\,$\mu$m	& 1.484 $\pm$\ 0.009	& $-$0.763 $\pm$\ 0.009	& 6.716	& 0.13	\\ 
F\,37.1\,$\mu$m	& F\,25.3\,$\mu$m	& 1.409 $\pm$\ 0.009	& $-$0.687 $\pm$\ 0.008	& 6.754	& 0.12	\\ 
F\,34.8\,$\mu$m	& F\,19.7\,$\mu$m	& 1.058 $\pm$\ 0.006	& $-$0.361 $\pm$\ 0.005	& 6.590	& 0.13	\\ 
F\,34.8\,$\mu$m	& F\,24.2\,$\mu$m	& 1.62 $\pm$\ 0.01		& $-$0.91 $\pm$\ 0.01	& 6.63	& 0.13	\\ 
F\,34.8\,$\mu$m	& F\,25.3\,$\mu$m	& 1.52 $\pm$\ 0.01		& $-$0.815 $\pm$\ 0.009	& 6.681	& 0.13	\\ 
F\,33.6\,$\mu$m	& F\,19.7\,$\mu$m	& 1.074 $\pm$\ 0.006	& $-$0.384 $\pm$\ 0.005	& 6.532	& 0.14	\\ 
F\,33.6\,$\mu$m	& F\,24.2\,$\mu$m	& 1.71 $\pm$\ 0.01		& $-$1.01 $\pm$\ 0.01	& 6.56	& 0.13	\\ 
F\,33.6\,$\mu$m	& F\,25.3\,$\mu$m	& 1.60 $\pm$\ 0.01		& $-$0.90 $\pm$\ 0.01	& 6.62	& 0.13	\\ 
F\,31.5\,$\mu$m	& F\,19.7\,$\mu$m	& 1.131 $\pm$\ 0.007	& $-$0.455 $\pm$\ 0.006	& 6.443	& 0.14	\\ 
F\,31.5\,$\mu$m	& F\,24.2\,$\mu$m	& 2.04 $\pm$\ 0.02		& $-$1.35 $\pm$\ 0.01	& 6.48	& 0.14	\\ 
F\,31.5\,$\mu$m	& F\,25.3\,$\mu$m	& 1.86 $\pm$\ 0.01		& $-$1.17 $\pm$\ 0.01	& 6.56	& 0.13	\\ 
\hline  \\[-9pt]
\end{tabular}
\end{center}
\footnotesize{{$^a$$\sigma_L$ represents the dispersion between best-fit and input $\log{L_{\rm int}}$ values, for models detected by FORCAST in both filters.}}\\
\end{table*}

Equation~\ref{singlefit3}, with fluxes at an adopted distance of 140~pc, may be converted to a form directly applicable to observations, for which $S_\nu$ is typically given in Jy and valid for any distance $d$, as
\begin{equation}
L_{\rm int}{\rm (fit)} = \left[ \left( \frac{d}{140\,{\rm pc}} \right)^2 \frac{\nu S_{\nu} }{10^{23+b}} \right]^{1/m} L_\odot
\end{equation}

\noindent
where $\nu$ is the effective frequency, given in Hz, of the filter, and the best-fit parameter values $m$ and $b$ may be obtained from Table~\ref{linearfits}.  Focusing on 70~$\mu$m, for example, $L_{\rm int}$ may be estimated using
\begin{equation}
L_{\rm int}{\rm (fit)} = 0.115 \left[ \left( \frac{d}{140\,{\rm pc}} \right)^2 S_{\nu,{\rm 70}}\right]^{0.855} L_\odot .
\label{singlefitfocus}
\end{equation}

\noindent
Thus, a protostar observed at 70~$\mu$m to be 1 Jy at a distance of 140~pc suggests that $L_{\rm int}$ is $\sim$0.1~$L_\odot$, reliable to within a factor of 2.3 (3$\sigma$), as previously discussed.

With the scatter in the correlations between FORCAST fluxes and $L_{\rm int}$ being primarily a function of inclination, we explored whether utilizing two FORCAST fluxes may improve estimates of $L_{\rm int}$.  Again referring to Figure~14 of \cite{whitney2013}, the slopes of the SEDs in the 20--40~$\mu$m regime appear to be correlated with inclination, suggesting that two FORCAST fluxes would in principle provide a first-order luminosity estimate from the average flux level and second-order correction to the estimate from the slope.
For example, \cite{kryukova2012} found that, for protostars lacking 70~$\mu$m fluxes, better luminosity estimates could be achieved by considering both the Spitzer 24~$\mu$m fluxes and slopes of available 3.6--24~$\mu$m SEDs than by considering only 24~$\mu$m fluxes alone.  Such luminosity estimates were reliable to within a factor of $\sim$11 (3$\sigma$), compared to a factor of 48 (3$\sigma$; from $\sigma_L = 0.56$ in Table~\ref{linearfits}) based on 24~$\mu$m fluxes alone, representing a marked improvement.

Our approach to use two FORCAST fluxes is similar to that by \cite{kryukova2012} to use Spitzer 3.6--24~$\mu$m SEDs, but we might expect a greater improvement since FORCAST extends to longer mid-infrared wavelengths. Toward this end, we considered pairs of FORCAST filters, where the first filter was one of longer wavelengths (i.e., 31.5--37.1~$\mu$m) and the second filter was one of shorter wavelengths (i.e., 19.7--25.3~$\mu$m).  Linear regression was then used to determine the best-fit coefficients to
\begin{equation}
\log{L_{\rm int}} = C_1 \log{F_{\nu 1}} + C_2 \log{F_{\nu 2}} + C_3
\label{doublefit}
\end{equation}

\noindent
where the fluxes at 140~pc associated with Filter 1 and Filter 2 are denoted as $F_{\nu 1}$ and $F_{\nu 2}$, respectively.  Table~\ref{regressionfits} lists these coefficients for the different filter pairs, and Figure~\ref{LintEstimates} compares $L_{\rm int}{\rm (fit)}$ with those input into the model.  In general, there is reasonable agreement for all models, particularly those detectable by FORCAST, with increased dispersion for intrinsically fainter protostars.  Similar to Table~\ref{linearfits}, Table~\ref{regressionfits} also lists values for $\sigma_L$ to quantify the reliability of luminosity estimates based on these fits for all models detectable by FORCAST.

Regardless of the FORCAST filter combination, the two-filter fits provide luminosity estimates reliable at least to within a factor of 2.6 (3$\sigma$).  Filter combinations including F\,25.3\,$\mu$m generally provide the most constrained estimates.  The best FORCAST filter combination is F\,37.1\,$\mu$m with F\,25.3\,$\mu$m, which yields $\sigma_L = 0.12$, providing luminosities reliable to within a factor of 2.3 (3$\sigma$), comparable to that achieved from 70~$\mu$m observations.

Our fits to Equation~\ref{doublefit} may be recast in a form more directly applicable to observations of a source at distance $d$, in general, as
\begin{equation}
L_{\rm int}{\rm (fit)} = \Lambda \left( \frac{d}{140\,{\rm pc}} \right)^{2(C_1+C_2)} S_{\nu1}^{\,C_1} \,S_{\nu2}^{\,C_2} L_\odot ,
\label{doublefitobs}
\end{equation}

\noindent
where $S_{\nu 1}$ and $S_{\nu 2}$ are the observed flux densities in Jy in Filters 1 and 2, respectively, and $\Lambda$ is the coefficient accounting for the conversion of units and overall normalization given by
\begin{equation}
\Lambda = \frac{\nu_1^{C_1} \nu_2^{C_2}}{10^{23(C_1+C_2)-C_3}} ,
\label{gamma}
\end{equation}

\noindent
where $\nu_1$ and $\nu_2$ are the effective frequencies, in Hz, associated with Filters 1 and 2, respectively.  For example, focusing explicitly on 37.1~$\mu$m and 25.3~$\mu$m, $L_{\rm int}$ may be estimated using
\begin{eqnarray}
L_{\rm int}{\rm (fit)} = & 0.226 \left( \frac{d}{140\,{\rm pc}} \right)^{1.444} \left(S_{\nu,{\rm 37.1}}\right)^{1.409} \nonumber \\
& \times \left(S_{\nu,{\rm 25.3}}\right)^{-0.687} L_\odot ,
\label{Eqn3725Fit}
\end{eqnarray}

\noindent
where the flux densities $S_{\nu,{\rm 37.1}}$ and $S_{\nu,{\rm 25.3}}$ are given in Jy.

While FORCAST filter pairs yield $L_{\rm int}$ estimates with reliability comparable to those achieved previously with 70~$\mu$m observations, observational biases are evident in Figure~\ref{LintEstimates} and depend on the specific filter pair, protostellar luminosity, and sensitivity of the FORCAST observations.  Brighter protostars are preferentially detected, resulting in fitted luminosity estimates that systematically overestimate $L_{\rm int}$, especially evident for low luminosity protostars $L_{\rm int} < 0.3~$L$_\odot$ observed with a filter pair that includes F\,19.7\,$\mu$m.  The filter combination F\,37.1\,$\mu$m with F\,25.3\,$\mu$m shows the least bias, though the luminosities are still overestimated for the lowest luminosity protostars.  The degree to which $L_{\rm int}$ estimates are biased increase for relatively low luminosity protostars and for less sensitive observations.

Figure~\ref{LintEstimates} also shows that nearly all FORCAST-detectable models lie within the $3\sigma_L$ ranges, extrapolated from computed $\sigma_L$ dispersions listed in Table~\ref{regressionfits}, suggesting that these ranges overestimate the ranges associated with 99\% of FORCAST-detectable models.  For example, a careful analysis that accounts for the asymmetric and non-normal distribution suggests that $L_{\rm int}$(fit) is consistent with $L_{\rm int}$ to within a factor of 1.9--2.1 for 99\% of models detectable by F\,37.1\,$\mu$m and F\,25.3\,$\mu$m, slightly smaller than the factor of 2.3 extrapolated from $\sigma_L$.  (A difference in $\sigma_L$ of only 0.01--0.02 accounts for this effect.)  In other words, $\sigma_L$ slightly {\it{understates}} the reliability of $L_{\rm int}$ estimates derived from FORCAST filter pairs.  Given different systematic uncertainties, such as discussed in \S\ref{apsizes} and \S\ref{dustimpact}, we continue to adopt $\sigma_L$ from Table~\ref{regressionfits} as they are conservative measures of the reliability of $L_{\rm int}$ estimates.

\section{DISCUSSION}\label{discussion}

\begin{figure*}[t!]
\begin{center}
\includegraphics[trim=0.55in 1.3in 0.5in 0in, width=8.1in]{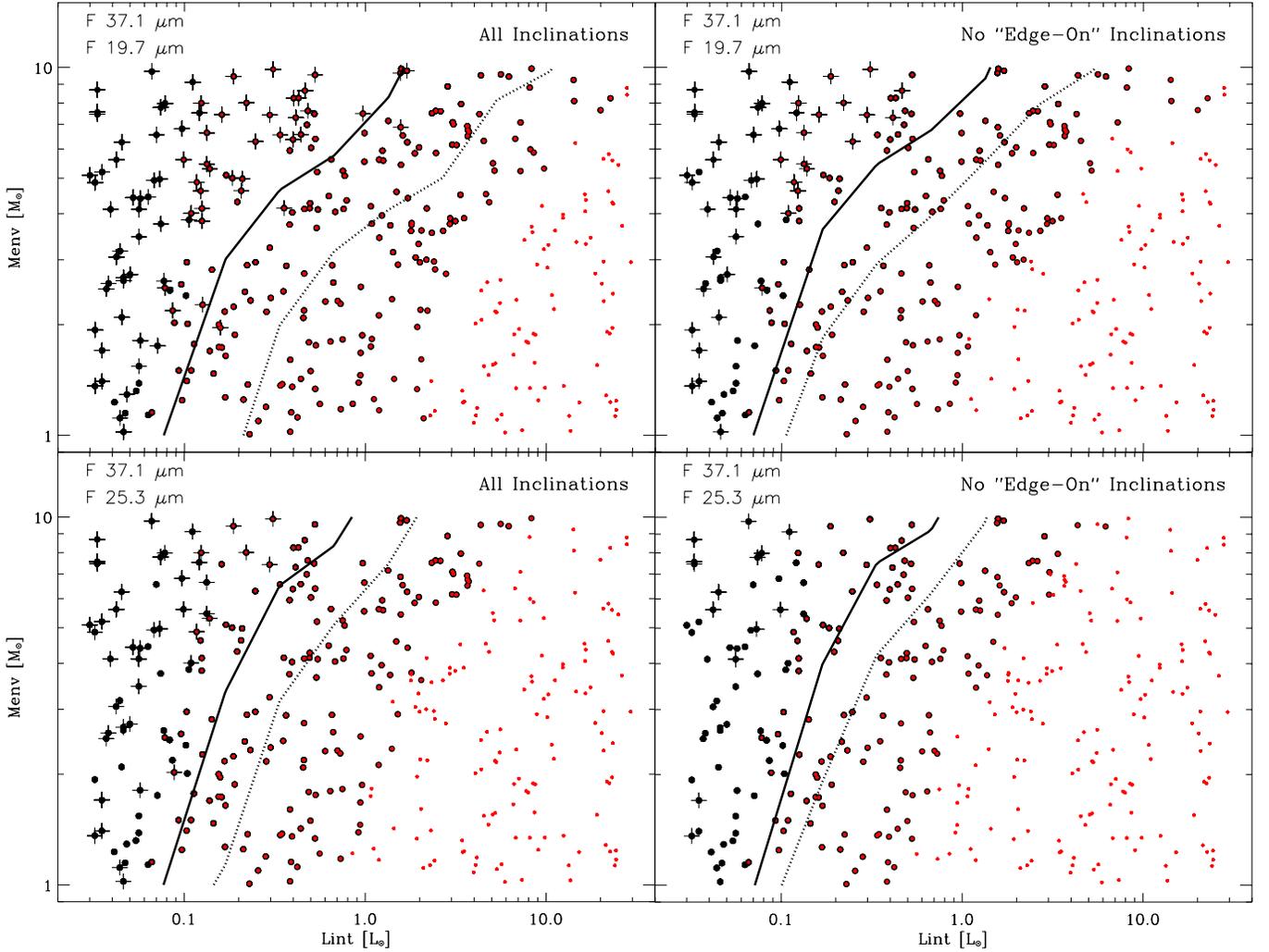}
\end{center}
\caption{Comparison of ($L_{\rm int}$, $M_{\rm env}$) parameter space probed by our models and those detectable by FORCAST, for select pairs of filters and inclinations.  Top panels are models observed with filters F\,37.1\,$\mu$m and F\,19.7\,$\mu$m; bottom panels are models observed with filters F\,37.1\,$\mu$m and F\,25.3\,$\mu$m.  Left panels include models for all inclinations, while the right panels exclude models with nearly edge-on inclinations (specifically, with $\cos{i} < 0.2$).  First, models that are undetectable by FORCAST are plotted with larger and thicker black symbols, and then models that are detectable by FORCAST are overplotted as smaller and thinner red symbols.  The symbol is a dot for a model dominated by protostellar radiation; it is a plus sign for a model dominated by scattered or reprocessed radiation from the ISRF.  Note that there are no red plus signs plotted since all models dominated by ISRF are undetectable by FORCAST and therefore appear as black plus signs.  Since there are 10 models (one for each inclination) for each ($L_{\rm int}$, $M_{\rm env}$) probed, only red dots appear for those models detectable for all inclinations; red dots on top of larger black dots appear for models detectable for only some inclinations; and only larger black dots appear for models undetectable for all inclinations.  Finally, the dotted and solid curves correspond to the region where 50\% and 25\%, respectively, of the models are detectable by FORCAST.
\label{discussionpanels}} 
\end{figure*}

As illustrated in Figures~\ref{MIPSFORCASTpanels} and \ref{LintEstimates}, FORCAST is not sensitive to the faintest of protostars.  While protostars at 140 pc with $L_{\rm int} \gtsim$ 0.2 L$_\odot$ are detectable at 37.1~$\mu$m, only those more luminous than $\sim$0.7~L$_\odot$ are detectable at 19.7~$\mu$m.  Our method to use a pair of FORCAST filters to determine internal luminosities of protostars is viable for protostars detectable in those filters, which is driven primarily by the sensitivity at shorter wavelengths.  Thus, the choice of filter pair is important.  Furthermore, the viability of this method is not solely a function of the internal luminosity, but inclination and other properties play a role.

In Figure~\ref{discussionpanels}, we explore the interplay of internal luminosity, envelope mass, and inclination in determining the detectability of protostars with two different filter pairs: F\,37.1\,$\mu$m with F\,19.7\,$\mu$m; and F\,37.1\,$\mu$m with F\,25.3\,$\mu$m.  Comparing the two left panels, we see that protostars of lower luminosities are detectable when using F\,25.3\,$\mu$m rather than F\,19.7\,$\mu$m, as expected, especially for greater envelope masses.  While envelope mass affects detectability, it has less impact (i.e., the solid and dotted curves exhibit steeper slopes) when using F\,25.3\,$\mu$m.  Comparing each of the right panels with its adjacent left panel, we see that a greater fraction of lower luminosity protostars are detectable, if those models with nearly edge-on protostars are excluded.

The sensitivity of F\,25.3\,$\mu$m, over that of F\,19.7\,$\mu$m, to more models is a compelling reason to favor it.  As previously mentioned, Table~\ref{regressionfits} demonstrates that F\,25.3\,$\mu$m paired with F\,37.1\,$\mu$m yields $L_{\rm int}$ estimates with less uncertainty.  Given these considerations of completeness and precision, observations of protostars with F\,37.1\,$\mu$m and F\,25.3\,$\mu$m are likely best for the purpose of determining $L_{\rm int}$.

We find models for which more than half of the radiation within 20\arcsec-radius (2800 AU at 140 pc) apertures are reprocessed or scattered photons originating from the external ISRF rather than from the protostellar system.  While such observed ISRF photons are dependent on the strength of field, extinction from the parental molecular cloud, and properties of the protostellar envelope, they are not tied to the internal protostellar luminosity.  Thus, contribution (or ``contamination'') from the ISRF primarily serves to add scatter in the relationships between $F_{\nu}$ and $L_{\rm int}$, and it enables a guide to the level of precision possible on estimates of $L_{\rm int}$ for protostellar systems in a typical range of environments.  ISRF-dominated models are more prevalent at 19.7~$\mu$m than at 25.3~$\mu$m and are found more for nearly edge-on, less luminous protostars.  While \textsc{Hochunk3d} enables tracking of the sources of the photons imaged from each system, an observer, in general, does not know {\it{a priori}} the relative contribution of the ISRF.  It may be possible to estimate this contamination based on the observed radial profile, enabling results with better $L_{\rm int}$ precision.  For our study, we did not pursue such an investigation since we were able to estimate $L_{\rm int}$ with comparable precision to that provided by Spitzer and Herschel.  Furthermore, models dominated by the ISRF in the mid-infrared FORCAST bands are at least two orders of magnitude below the fiducial FORCAST sensitivity limits.

\subsection{Inclination and External Heating}\label{flashlight}

Traditionally, the luminosity of a protostar has been determined by integrating an SED, which requires sufficient spectral coverage, especially from the infrared to submillimeter regimes.  We refer to this luminosity as the {\it{observed}} bolometric luminosity, $L_{\rm bol}{\rm (SED)}$.  The evacuated cavity and protostellar disk primarily, and envelope density profile secondarily, result in a non-uniform escape of infrared photons.  Light detected from a pole-on protostar suffers less extinction relative to the same protostar observed edge-on.  Therefore, $L_{\rm bol}{\rm (SED)}$ will overestimate the {\it{true}} bolometric luminosity in the case of the pole-on protostar and underestimate it for the edge-on protostar. This effect, which we refer to as the ``flashlight effect,'' has been documented in the literature (e.g., \citealt{zhang2013}; \citealt{whitney2003a}; \citealt{yorke1999}).  For example, in their Figure~10, \cite{whitney2003a} demonstrated that $L_{\rm bol}{\rm (SED)}$ overestimated the true bolometric luminosity by about a factor of 2 for their pole-on protostars with bipolar cavities and underestimated it by 50\% for the same protostars observed edge-on.  Our models show a similar trend.

The bolometric luminosity includes the internal luminosity of the protostar as well as a component, or ``contamination,'' due to external heating by the ISRF.  For our models, this contamination is typically $\sim$0.3~$L_\odot$, consistent with previous studies (e.g., \citealt{evans2001}), and reaches as high as $\sim$1~$L_\odot$ in some cases.  Thus, not only does $L_{\rm bol}{\rm (SED)}$ suffer from the flashlight effect, but the contamination from external heating can be significant, particularly for protostars with $L_{\rm int} \ltsim$~1~$L_\odot$.

\begin{figure}[t!]
\begin{center}
\begin{turn}{90}
\includegraphics[trim=1.2in 0in 0.6in 1.65in, height=3.5in]{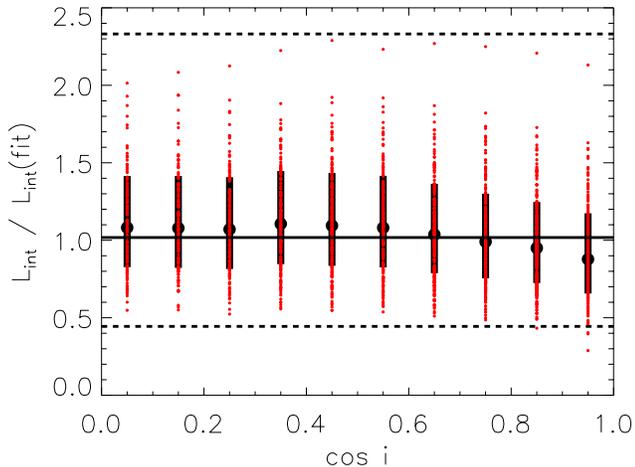}
\end{turn}
\end{center}
\caption{Plot of $L_{\rm int} / L_{\rm int}{\rm (fit)}$ as a function of $\cos{i}$ for models detectable by FORCAST filters F\,37.1\,$\mu$m and F\,25.3\,$\mu$m, demonstrating that our method using this filter pair yields $L_{\rm int}$ estimates that do not depend on inclination.  Individual models are represented by red points (the same models plotted by red points in the bottom right panel of Figure~\ref{LintEstimates}), while the median and dispersion (computed in log space) at each inclination are represented by the black filled circles and error bars.  The horizontal solid line represents the median value for all detectable models, and dashed lines represent the 3$\sigma_L$ range.  Plots for other filter pairs show similar results.
\label{LintINC}} 
\end{figure}

In our method to estimate protostellar luminosities from FORCAST fluxes, the fluxes were empirically fit to $L_{\rm int}$ (Equation~\ref{doublefit}); thus, it calibrates out the effect of external heating, in a statistical sense.  But, does our method suffer from the flashlight effect?  Utilizing a pair of FORCAST filters was intended to account for inclination, the primary factor in the large scatter in correlations between FORCAST fluxes and $L_{\rm int}$ shown in Figure~\ref{MIPSFORCASTpanels}.  In Figure~\ref{LintINC}, we plot $L_{\rm int}$ relative to $L_{\rm int}$(fit) derived from F\,37.1\,$\mu$m and F\,25.3\,$\mu$m, as a function of inclination, demonstrating that our method results in reliable luminosity estimates that do not depend on inclination.  Plots for other filter pairs show similar results.  Thus, our method successfully uses pairs of FORCAST filters to estimate 
$L_{\rm int}$ to better characterize protostars more efficiently than obtaining a full SED to determine $L_{\rm bol}{\rm (SED)}$.

\subsection{Consideration of Aperture Sizes}\label{apsizes}

Deriving FORCAST fluxes from 6\arcsec-radius apertures enabled us to compare our results directly with \citealt{dunham2008}, who used the same aperture size for 10--40~$\mu$m.  In principle, the flux-luminosity relationships derived by our study and by \citealt{dunham2008} apply strictly to fluxes derived with the same physical size of the aperture -- i.e., radius of 840~AU.  In practice, however, because these apertures include most (typically $\gtsim$90\%) of the mid-infrared emission from the protostars, it is not important to adhere to the same physical aperture size when deriving fluxes for protostars at different distances, if the apertures include a larger physical area.  For example, one could simply use 6\arcsec-radius apertures to measure fluxes for all protostars in the Gould's Belt molecular clouds, with distances $\sim$140--500~pc.  The relatively small amount of flux added by including a region of 3000~AU for protostars at 500~pc compared to a region of 840~AU for those at 140~pc potentially introduces a systematic error that is insignificant compared to other dominant systematic errors.  Furthermore, since our relationships involve pairs of FORCAST filters, such a systematic error is expected to be even more muted since the measured mid-infrared FORCAST fluxes will be increased similarly in both filters when increasing the physical aperture size.  

We tested these expectations explicitly by using total flux densities (obtained from large 100\arcsec-radius apertures capturing emission from the entire protostellar envelope of outer radius 14,000~AU) in the relationships given by Equation~\ref{doublefitobs} to derive $L_{\rm int}$ estimates that were typically within 5\% of those obtained from the 6\arcsec-radius apertures from which the relationships were derived.  We do not recommend using Equation~\ref{doublefitobs} (or Equation~\ref{Eqn3725Fit}) with flux densities obtained from apertures much smaller than 840~AU since the more extended size of the protostar (from scattered light) at shorter FORCAST wavelengths relative to that at longer wavelengths may result in greater systematic errors. 

\subsection{Impact of Dust Grain Population}\label{dustimpact}

While OH5 grain opacities are most consistent with observational constraints, there are discrepancies, as discussed in \S\ref{dustsection}.  To quantify the impact of the shortcoming of OH5 grains on $L_{\rm int}$ estimates derived from a pair of FORCAST filters, we reran our protostellar models using ``revised OH5'' grains, with 35\% greater opacity for $\lambda \ge 2.5~\mu$m compared to OH5 grains.  The flux densities at 37.1~$\mu$m and 25.3~$\mu$m for all FORCAST-detectable models were used to obtained $L_{\rm int}$ estimates using Equation~\ref{Eqn3725Fit}.  These estimates were typically $\sim$5\% less than those obtained using OH5 grains; most models with revised OH5 grains yielded $L_{\rm int}$ estimates that were within 5--10\% of those obtained with OH5 grains.  We therefore expect any improved dust model to have a relatively small effect on our results.

\subsection{Applicability of our Results}

We stress that our results apply to those embedded protostellar sources at an early evolutionary stage exhibiting a protostellar envelope, as described in \S\ref{models}.  Historically, such protostars have been observationally identified as Class~0 or Class~I (hereafter, Class~0/I) sources, based on thermal dust emission or the slopes, $\alpha$, of the infrared SEDs (\citealt{lada1987}, \citealt{andre1993}).  Class~0/I sources are commonly defined as those with $\alpha \geq$ 0.3, while Flat sources are those with 0.3 $> \alpha \geq$ $-$0.3 and Class~II sources, primarily representing evolved YSOs, are those with $-$0.3 $> \alpha\ \geq$ $-$1.6 (\citealt{greene1994}).

While sources observationally identified as Class~0/I are likely bona fide protostars with envelopes, it is possible that some of these sources instead are the more evolved Class~II sources obscured by sufficient molecular cloud material such that their SEDs mimic those of protostars.  Such ``contamination'' is most prevalent in embedded young clusters, where the intracluster material may provide significant extinction along the lines of sight.  Based on a Spitzer study of the NGC~2264 and IC~348 clusters (\citealt{forbrich2010}), up to a third of sources previously identified as Class~0/I were found to be consistent with extincted Class~II sources, though half of these possible extincted Class~II sources are also still consistent with being Class~0/I protostars.  More recently, \cite{carney2016} use HCO$^+$ J=3--2, C$^{18}$O J=3--2, and 850~$\mu$m observations to distinguish Class 0/I protostars from extincted Class~II sources in Perseus and Taurus; they found $\sim$30\% of sources classified as Class~0/I based on their SEDs were likely extincted Class~II sources. Thus, most sources classified as Class~0/I based on their infrared SEDs are bona fide Class~0/I protostars, especially in relatively isolated star-forming regions, with no more than $\sim$20--30\% expected to be extincted Class~II sources in regions of embedded clusters.  If a source exhibits $\alpha \geq$ 0.3, it is most likely a protostar, and our results most likely apply.

\begin{figure}[t!]
\begin{center}
\includegraphics[trim=0.25in 0in 0.2in 0in, width=3.2in]{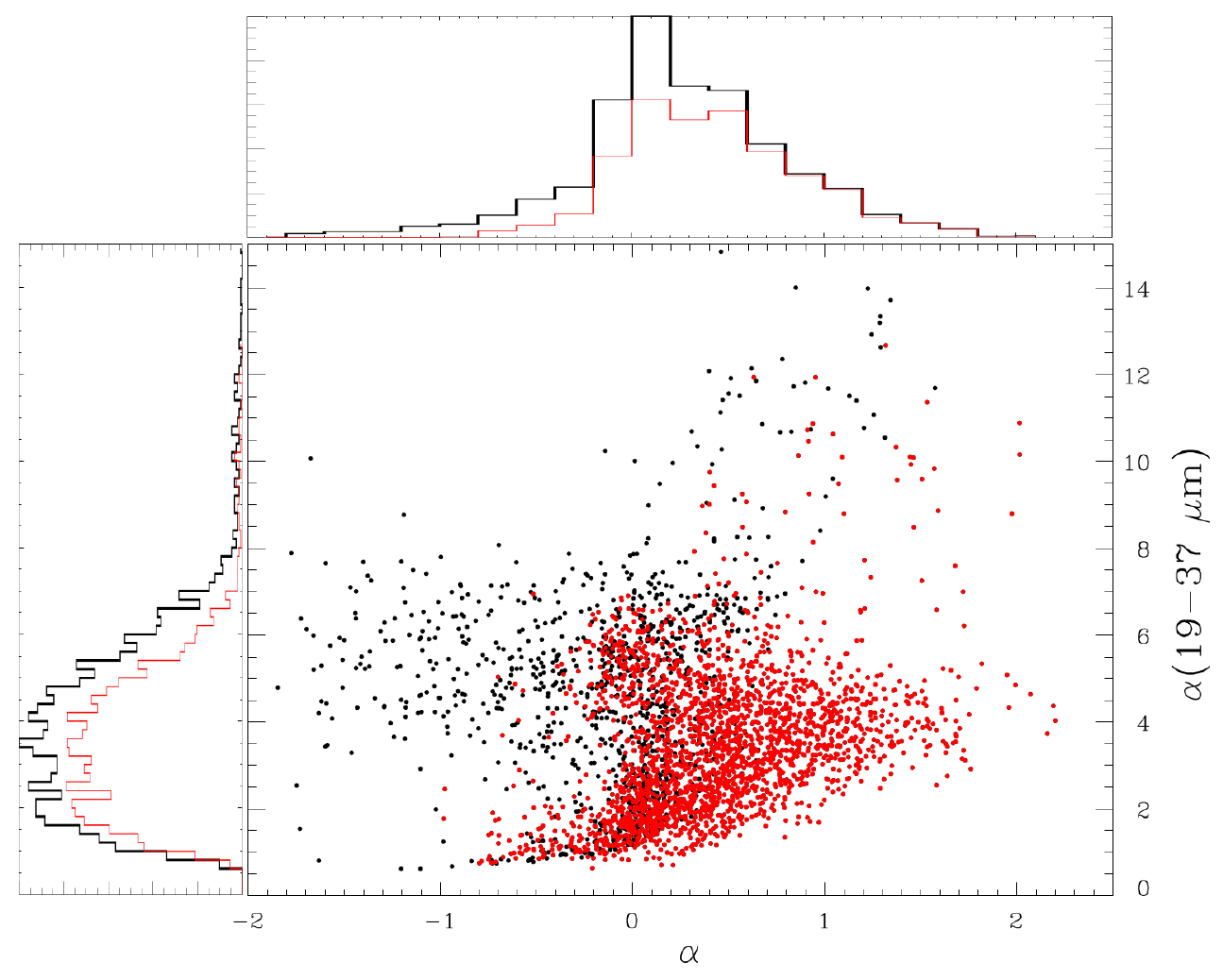}
\end{center}
\caption{Slopes of the SEDs characteristic of our protostellar models.  The mid-infrared slopes $\alpha\ (19$--$37~\mu m)$, derived from FORCAST bands, are plotted relative to the infrared slopes $\alpha$ traditionally defined by the 2--24~$\mu$m bands.  Those models detectable by FORCAST are plotted as red points, while those models not detectable are plotted as black points.  The histograms along the left represent the distributions of $\alpha\ (19$--$37~\mu m)$, with black showing the distribution for all models and red showing that for only the detectable models.  The histograms along the top represent the distributions of $\alpha$, with the same color scheme. Focusing on the FORCAST-detectable models, the median slopes are $\alpha \sim\ 0.3$ and $\alpha(19$--$37~\mu m) \sim\ 3.5$, with 95\% of these models falling into the following ranges: $\alpha= [-0.3, 0.2]$ and $\alpha(19$--$37~\mu m) = [1, 6.5]$.
\label{composite}} 
\end{figure}

A source exhibiting $\alpha <\ $0.3 may also be a protostar described by our modeling; in fact, the distribution of $\alpha$ exhibited by our models, shown in Figure~\ref{composite}, peaks at $\alpha \sim\ $0.1 and shows an extended tail for $\alpha\ \leq\  -0.3$, slopes that are more indicative of Flat-spectrum and Class II sources. Since most decreasing infrared SEDs are associated with Class II sources, additional evidence beyond the 2--24~$\mu$m SED would be necessary to believe reasonably that a particular source with such an SED is a protostar, rather than an evolved Class II source.

The case for identifying a source as a protostar may be bolstered by FORCAST 19--37~$\mu$m observations.  Figure~\ref{composite} shows that {\it{all}} protostellar models exhibit mid-infrared SED slopes, $\alpha(19$--$37~\mu m) > 0.5$, in the FORCAST bands.  Our distribution of $\alpha(19$--$37~\mu m)$ is consistent with previously published SEDs.  For example, inspection of the SEDs of a standard Class~I protostar from Whitney et al. (2013; Figure 14) suggests that $\alpha(19$--$37~\mu m)\ \gtsim\ 1$ for all inclinations; earlier stage Class~0 protostars would exhibit greater values. Furthermore, inspection of the SEDs of Class~II sources from Whitney et al. (2013; Figure 2) suggest flatter mid-infrared SED slopes, with $\alpha(19$--$37~\mu m)\ \ltsim\ 0.5$ for most inclinations; only relatively edge-on inclinations with $\cos{i}\ \ltsim\ 0.25$ have $\alpha(19$--$37~\mu m)$ that rival those of protostars.

The identification and classification of protostars is beyond the scope of this paper, but we have provided some considerations based on the traditionally defined $\alpha$ and how FORCAST observations to derive $\alpha(19$--$37~\mu m)$ may help.  If the sources are indeed protostars, those same FORCAST observations can be used to estimate their internal luminosities by Equation~\ref{doublefitobs} (or, Equation~\ref{Eqn3725Fit} specifically for 25.3~$\mu$m and 37.1~$\mu$m observations).

\subsection{Envelope Mass}\label{envsection}

The assumptions and applicability of the Ulrich envelope density profile are important to consider, especially
in the context of high envelope masses relevant for protostar models.  The assumptions include: 1) free-fall toward a central mass, $M_* + M_{\rm disk}$, at the free-fall velocity given by
\begin{equation}
\vff = \sqrt{\frac{2 G (M_*+ M_{\rm disk})}{R_{\rm env}}} ;
\label{VffEQN}
\end{equation}

\noindent
setting $r = R_{\rm env}$ to focus on effects near the adopted cloud boundary and 2) pressure terms that are small compared to kinetic energy, which can be written as the condition
\begin{equation}
\vff \ge a_s ,
\label{lowpressureEQN}
\end{equation}

\noindent
where $a_s$ is the thermal sound speed of the gas.  For example, the fiducial case of $R_{\rm env} = 14,000$ {\rm AU} and $M_* + M_{\rm disk} \approx M_* = 0.5~{\rm M_\sun}$, as assumed in our modeling, results in $\vff = 0.25~{\rm km~s}^{-1}$, which is on order of the thermal sound speed of $a_s = 0.19  ~{\rm km~s}^{-1}$ for $T = 10 ~{\rm K}$ gas (e.g., \citealt{terebey1984}).  Thus, pressure terms are important near the adopted edge of the envelope, with the result that the density distributions in real protostellar envelopes will deviate from that given by the Ulrich profile near the envelope boundaries.

The first assumption, as expressed in Equation~\ref{VffEQN}, applies if the central mass dominates the gravitational potential.  However, the envelope masses considered here and in previous studies (e.g., \citealt{dunham2008}), which also use the Ulrich envelope model, typically exceed the central masses.  To evaluate the size of the effect, we compare the Ulrich envelope mass computed from Equation~\ref{MenvEQN}, which further assumes the central mass is dominated by the star (i.e., $M_{\rm disk} \ll M_*$), with that of the TSC84 (\citealt{terebey1984}) self-consistent cloud collapse model.  This comparison is appropriate because the TSC84 model includes pressure effects and asymptotically matches the Ulrich model at small radii.

\begin{figure}[t!]
\begin{center}
\includegraphics[trim=0.2in 0.1in 0in 0in, width=3.35in]{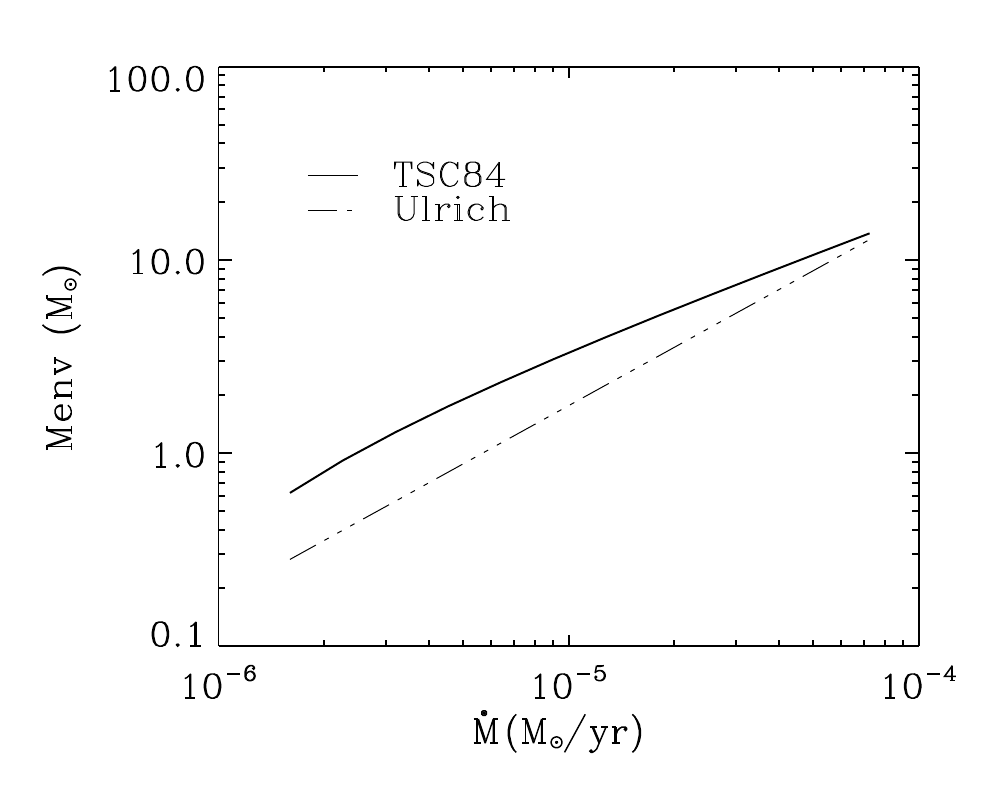}
\includegraphics[trim=0.2in 0.2in 0in 0in, width=3.35in]{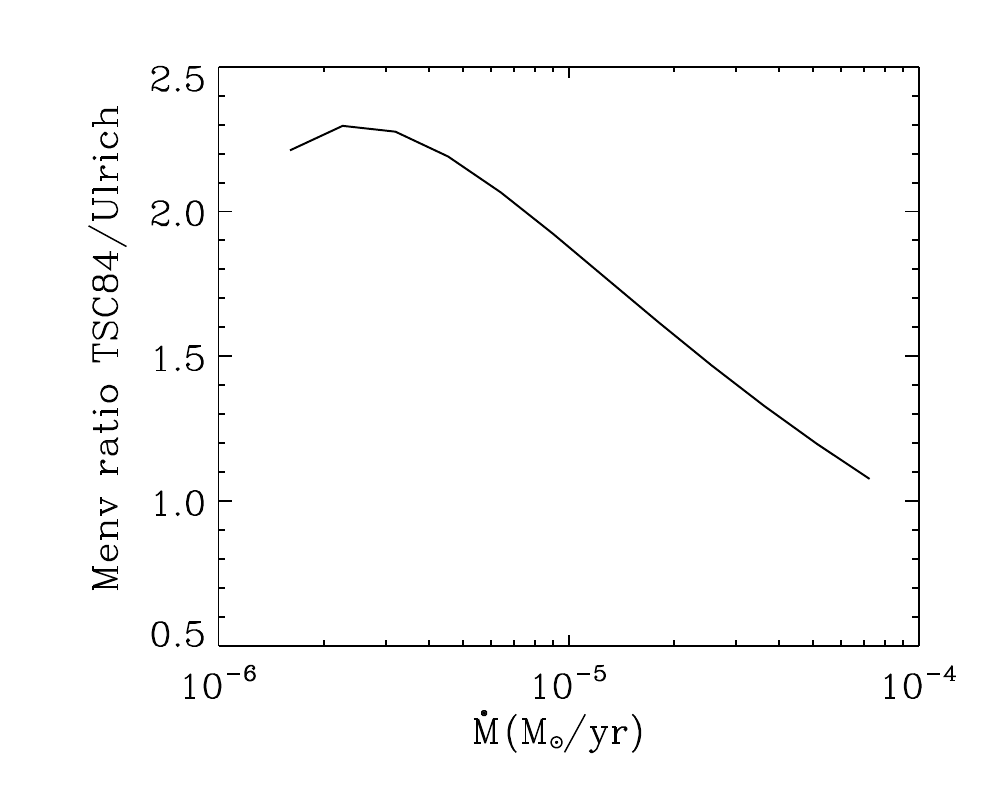}
\end{center}
\caption{Comparison of $M_{\rm env}$ (top) and $M_{\rm env}$ ratio (bottom) between the TSC84 and Ulrich solutions as a function of mass infall rate, $\mdotenv$, for $M_* = 0.5 M_\odot$.
The Ulrich solution underestimates the envelope mass for the protostar case. 
\label{menvfactor}} 
\end{figure}

In the TSC84 model, {\it outside} the collapsing region of the envelope {\it and} when rotational effects are small, a simple formula (involving the leading term) gives the total mass interior to $r$ (e.g., Equation~3 of TSC84; \citealt{shu1977}):
\begin{equation}
M_{{\rm tot}} = \frac{2 a_s^2 r} {G},
\end{equation}

\noindent
where $r > R_{\rm exp}(t)$, the radius of the expansion wave representing the boundary of the collapsing region at
time $t$ and given by $R_{\rm exp}(t) = a_s t$.  In terms of the mass infall rate, given by
\begin{equation}
\dot{M}_{\rm env} = \frac{m_0 a_s^3}{G} , 
\end{equation}

\noindent
where $m_0 = 0.975$ (\citealt{shu1977}), this total mass may be expressed as
\begin{equation}
M_{{\rm tot}} = \left( \frac{2}{m_0} \right) \frac { \dot{M}_{\rm env} r} {a_s}.
\label{MtotTSC}
\end{equation}

\noindent
The difference between this total mass and the central mass, which has already collapsed to form the protostar and disk, is the desired envelope mass interior to $r > R_{\rm exp}$ in the TSC84 model:  
\begin{equation}
M_{\rm env}^{\rm TSC84} = M_{\rm tot} - (M_* + M_{\rm disk}) \\
\end{equation}

\noindent
Using Equation~\ref{MtotTSC} to substitute for $M_{\rm tot}$, and noting that the central mass is
\begin{equation}
M_* + M_{\rm disk} = \dot{M}_{\rm env} t ,
\end{equation}

\noindent
the envelope mass interior to $r > R_{\rm exp}$ is then given by
\begin{equation}
M_{\rm env}^{\rm TSC84} = \left( \frac {2}{m_0} - \frac {R_{\rm exp}}{r} \right) \frac{\dot{M}_{\rm env} r} {a_s} .
\label{MenvTSC}
\end{equation}

\noindent
The factor in parenthesis has a value of order unity at $r = R_{\rm exp}$ and of order 2 at $r \gg R_{\rm exp}$. Thus, the factor in parenthesis ranges from about 1 to 2 over the valid $r \ge R_{\rm exp}$ regime.  Note that computing the envelope mass at smaller $r$ requires using the full collapse solution, and can be done numerically, but it is not necessary for our purpose.

For direct comparison with the envelope mass in the Ulrich free-fall model, we rewrite Equation~\ref{MenvEQN} in terms of $\vff$, using Equation~\ref{VffEQN}, to obtain
\begin{eqnarray}
M_{{\rm env}} &= &\left( \frac{2 a_s}{3 \vff} \right) \frac{\dot{M}_{\rm env} R_{\rm env}}{a_s}, 
\label{menv_ff}
\end{eqnarray}

\noindent
which becomes the inequality,
\begin{eqnarray}
M_{{\rm env}} &\le & \frac{2}{3} \frac{\dot{M}_{\rm env} R_{\rm env}}{a_s} ,
\end{eqnarray}

\noindent
if indeed the pressure terms are small compared to kinetic energy (i.e., Equation~\ref{lowpressureEQN}).  This expression is similar in form to that of the TSC84 model in Equation~\ref{MenvTSC} above and demonstrates that the Ulrich profile underestimates the envelope mass, compared with a realistic envelope model having both gravity and pressure terms.  For the adopted case of $M_* = 0.5~M_\sun$, Figure \ref{menvfactor} compares the envelope mass at different accretion rates and at fixed $r = R_{\rm env} = 14,000 ~AU$ envelope radius. The mass difference is about a factor of two, large enough to be important at millimeter wavelengths where the observations are sensitive to cloud (i.e. envelope) mass.  However, the mass difference is less important at the mid-infrared wavelengths relevant to this study, where the fluxes generated are not sensitive to the treatment of the outer boundary (e.g. \citealt{whitney1993}).

\section{SUMMARY}\label{summary}

In this study, we have established an approach whereby a pair of FORCAST filters may be used to estimate the luminosities of protostars.  Empirical relationships are derived for different combinations of a long-wavelength filter (F\,31.5\,$\mu$m, F\,33.6\,$\mu$m, F\,34.8\,$\mu$m, F\,37.1\,$\mu$m) paired with a short-wavelength filter (F\,19.7\,$\mu$m, F\,24.2\,$\mu$m, F\,25.3\,$\mu$m).  We find that the best pairing is F\,37.1\,$\mu$m with F\,25.3\,$\mu$m, resulting in luminosity estimates reliable to within a factor of 2.3 for 99\% of protostars, which is comparable to the precision achievable in previous studies utilizing Spitzer or Herschel 70-$\mu$m data. The luminosity is estimated using Equation~\ref{Eqn3725Fit}
once the flux densities $S_{\nu,{\rm 37.1}}$ and $S_{\nu,{\rm 25.3}}$ in Jy are known. Table~\ref{regressionfits} gives results for other FORCAST filter pairs, which may be used with Equations~\ref{doublefitobs} and \ref{gamma} to estimate luminosities.

With many protostars lacking data at wavelengths 70~$\mu$m or longer, obtaining FORCAST observations and applying our results may be the best approach currently to determine their luminosities.  Furthermore, the higher angular resolution achievable by FORCAST enables partitioning of emission among sources blended in previous observations and better constraints on the SEDs and luminosities of the components.  Our approach requires data using only a pair of FORCAST filters, not a well covered SED in the infrared and submillimeter regimes, and is independent of the inclination of the protostar.

In \S\ref{dustsection} we consider available dust model opacities. Figure~\ref{obsgrainprops} shows that the OH5 opacity (augmented with Pollack optical constants) fits the observational data best, though not perfectly. We find that an improved dust model would affect the luminosity estimates by only 5--10\%, thus supporting the choice of OH5 dust for protostellar envelopes.

We also compare (\S\ref{envsection}) the commonly assumed Ulrich density profile for protostellar envelopes with a more realistic profile for envelope masses comparable to, or greater than, the embedded source.  Real protostellar envelopes likely have material collapsing slower than assumed free fall velocities, and pressure terms become appreciable in the outer regions of the envelopes.  Such considerations suggest that the Ulrich profile underestimates total envelope masses by about a factor of two, but this deficiency has little effect on observed infrared emission from the protostar.

\acknowledgments

We thank Joe Adams for providing FORCAST transmission curves for typical observing conditions, Michael Dunham for providing properties of OH5 dust, and Barbara Whitney for discussion of implementation of the external ISRF in \textsc{Hochunk3d}.  S. Terebey thanks the Department of Astronomy at  the University of Maryland for its hospitality.  Simulations were performed on the YORP cluster administered by the Center for Theory and Computation, part of the Department of Astronomy at the University of Maryland.  This study was based in part on observations made with the NASA/DLR Stratospheric Observatory for Infrared Astronomy (SOFIA).  SOFIA is jointly operated by the Universities Space Research Association, Inc. (USRA), under NASA contract NAS2-97001, and the Deutsches SOFIA Institut (DSI) under DLR contract 50 OK 0901 to the University of Stuttgart.  Financial support for this work was provided by NASA through awards SOF-0106 and SOF04-0173 issued through USRA.

\facility{SOFIA}
\software{HOCHUNK3D \citep{whitney2013}, SOFIA Instrument Time Estimator (\url{https://dcs.sofia.usra.edu/proposalDevelopment/SITE/})}

\bibliographystyle{aasjournal}
\bibliography{references}

\begin{thebibliography}{}
\expandafter\ifx\csname natexlab\endcsname\relax\def\natexlab#1{#1}\fi
\providecommand{\url}[1]{\href{#1}{#1}}

\bibitem[{{Adams} {et~al.}(2010){Adams}, {Herter}, {Gull}, {Schoenwald},
  {Henderson}, {Keller}, {De Buizer}, {Stacey}, \& {Nikola}}]{adams2010}
{Adams}, J.~D., {Herter}, T.~L., {Gull}, G.~E., {et~al.} 2010, in Society of
  Photo-Optical Instrumentation Engineers (SPIE) Conference Series, Vol. 7735,
  Society of Photo-Optical Instrumentation Engineers (SPIE) Conference Series,
  1

\bibitem[{{Andre} {et~al.}(1993){Andre}, {Ward-Thompson}, \&
  {Barsony}}]{andre1993}
{Andre}, P., {Ward-Thompson}, D., \& {Barsony}, M. 1993, \apj, 406, 122

\bibitem[{{Andr{\'e}} {et~al.}(2010){Andr{\'e}}, {Men'shchikov}, {Bontemps},
  {K{\"o}nyves}, {Motte}, {Schneider}, {Didelon}, {Minier}, {Saraceno},
  {Ward-Thompson}, {di Francesco}, {White}, {Molinari}, {Testi}, {Abergel},
  {Griffin}, {Henning}, {Royer}, {Mer{\'{\i}}n}, {Vavrek}, {Attard},
  {Arzoumanian}, {Wilson}, {Ade}, {Aussel}, {Baluteau}, {Benedettini},
  {Bernard}, {Blommaert}, {Cambr{\'e}sy}, {Cox}, {di Giorgio}, {Hargrave},
  {Hennemann}, {Huang}, {Kirk}, {Krause}, {Launhardt}, {Leeks}, {Le Pennec},
  {Li}, {Martin}, {Maury}, {Olofsson}, {Omont}, {Peretto}, {Pezzuto}, {Prusti},
  {Roussel}, {Russeil}, {Sauvage}, {Sibthorpe}, {Sicilia-Aguilar}, {Spinoglio},
  {Waelkens}, {Woodcraft}, \& {Zavagno}}]{andre2010}
{Andr{\'e}}, P., {Men'shchikov}, A., {Bontemps}, S., {et~al.} 2010, \aap, 518,
  L102

\bibitem[{{Bjorkman}(1997)}]{bjorkman1997}
{Bjorkman}, J.~E. 1997, in Lecture Notes in Physics, Berlin Springer Verlag,
  Vol. 497, Stellar Atmospheres: Theory and Observations, ed. J.~P. {De Greve},
  R.~{Blomme}, \& H.~{Hensberge}, 239

\bibitem[{{Black}(1994)}]{black1994}
{Black}, J.~H. 1994, in Astronomical Society of the Pacific Conference Series,
  Vol.~58, The First Symposium on the Infrared Cirrus and Diffuse Interstellar
  Clouds, ed. R.~M. {Cutri} \& W.~B. {Latter}, 355

\bibitem[{{Bolatto} {et~al.}(2007){Bolatto}, {Simon}, {Stanimirovi{\'c}}, {van
  Loon}, {Shah}, {Venn}, {Leroy}, {Sandstrom}, {Jackson}, {Israel}, {Li},
  {Staveley-Smith}, {Bot}, {Boulanger}, \& {Rubio}}]{bolatto2007}
{Bolatto}, A.~D., {Simon}, J.~D., {Stanimirovi{\'c}}, S., {et~al.} 2007, \apj,
  655, 212

\bibitem[{{Carlson} {et~al.}(2007){Carlson}, {Sabbi}, {Sirianni}, {Hora},
  {Nota}, {Meixner}, {Gallagher}, {Oey}, {Pasquali}, {Smith}, {Tosi}, \&
  {Walterbos}}]{carlson2007}
{Carlson}, L.~R., {Sabbi}, E., {Sirianni}, M., {et~al.} 2007, \apjl, 665, L109

\bibitem[{{Carlson} {et~al.}(2011){Carlson}, {Sewi{\l}o}, {Meixner}, {Romita},
  {Whitney}, {Hora}, {Cignoni}, {Sabbi}, {Nota}, {Sirianni}, {Smith}, {Gordon},
  {Babler}, {Bracker}, {Gallagher}, {Meade}, {Misselt}, {Pasquali}, \&
  {Shiao}}]{carlson2011}
{Carlson}, L.~R., {Sewi{\l}o}, M., {Meixner}, M., {et~al.} 2011, \apj, 730, 78

\bibitem[{{Carney} {et~al.}(2016){Carney}, {Y{\i}ld{\i}z}, {Mottram}, {van
  Dishoeck}, {Ramchandani}, \& {J{\o}rgensen}}]{carney2016}
{Carney}, M.~T., {Y{\i}ld{\i}z}, U.~A., {Mottram}, J.~C., {et~al.} 2016, \aap,
  586, A44

\bibitem[{{Cassen} \& {Moosman}(1981)}]{cassen1981}
{Cassen}, P., \& {Moosman}, A. 1981, \icarus, 48, 353

\bibitem[{{Chapman} {et~al.}(2009){Chapman}, {Mundy}, {Lai}, \&
  {Evans}}]{chapmanDUST2009}
{Chapman}, N.~L., {Mundy}, L.~G., {Lai}, S.-P., \& {Evans}, II, N.~J. 2009,
  \apj, 690, 496

\bibitem[{{Chapman} {et~al.}(2007){Chapman}, {Lai}, {Mundy}, {Evans}, {Brooke},
  {Cieza}, {Spiesman}, {Rebull}, {Stapelfeldt}, {Noriega-Crespo}, {Lanz},
  {Allen}, {Blake}, {Bourke}, {Harvey}, {Huard}, {J{\o}rgensen}, {Koerner},
  {Myers}, {Padgett}, {Sargent}, {Teuben}, {van Dishoeck}, {Wahhaj}, \&
  {Young}}]{chapman2007}
{Chapman}, N.~L., {Lai}, S.-P., {Mundy}, L.~G., {et~al.} 2007, \apj, 667, 288

\bibitem[{{Chiang} \& {Goldreich}(1997)}]{chiang1997}
{Chiang}, E.~I., \& {Goldreich}, P. 1997, \apj, 490, 368

\bibitem[{{Crapsi} {et~al.}(2008){Crapsi}, {van Dishoeck}, {Hogerheijde},
  {Pontoppidan}, \& {Dullemond}}]{crapsi2008}
{Crapsi}, A., {van Dishoeck}, E.~F., {Hogerheijde}, M.~R., {Pontoppidan},
  K.~M., \& {Dullemond}, C.~P. 2008, \aap, 486, 245

\bibitem[{{Draine}(1978)}]{draine1978}
{Draine}, B.~T. 1978, \apjs, 36, 595

\bibitem[{{Dullemond} \& {Dominik}(2004)}]{dullemond2004}
{Dullemond}, C.~P., \& {Dominik}, C. 2004, \aap, 417, 159

\bibitem[{{Dunham} {et~al.}(2008){Dunham}, {Crapsi}, {Evans}, {Bourke},
  {Huard}, {Myers}, \& {Kauffmann}}]{dunham2008}
{Dunham}, M.~M., {Crapsi}, A., {Evans}, II, N.~J., {et~al.} 2008, \apjs, 179,
  249

\bibitem[{{Dunham} {et~al.}(2010){Dunham}, {Evans}, {Terebey}, {Dullemond}, \&
  {Young}}]{dunham2010evolve}
{Dunham}, M.~M., {Evans}, II, N.~J., {Terebey}, S., {Dullemond}, C.~P., \&
  {Young}, C.~H. 2010, \apj, 710, 470

\bibitem[{{Dunham} {et~al.}(2014){Dunham}, {Stutz}, {Allen}, {Evans},
  {Fischer}, {Megeath}, {Myers}, {Offner}, {Poteet}, {Tobin}, \&
  {Vorobyov}}]{dunham2014}
{Dunham}, M.~M., {Stutz}, A.~M., {Allen}, L.~E., {et~al.} 2014, Protostars and
  Planets VI, 195

\bibitem[{{Dunham} {et~al.}(2015){Dunham}, {Allen}, {Evans},
  {Broekhoven-Fiene}, {Cieza}, {Di Francesco}, {Gutermuth}, {Harvey},
  {Hatchell}, {Heiderman}, {Huard}, {Johnstone}, {Kirk}, {Matthews}, {Miller},
  {Peterson}, \& {Young}}]{dunham2015}
{Dunham}, M.~M., {Allen}, L.~E., {Evans}, II, N.~J., {et~al.} 2015, \apjs, 220,
  11

\bibitem[{{Enoch} {et~al.}(2009){Enoch}, {Evans}, {Sargent}, \&
  {Glenn}}]{enoch2009}
{Enoch}, M.~L., {Evans}, II, N.~J., {Sargent}, A.~I., \& {Glenn}, J. 2009,
  \apj, 692, 973

\bibitem[{{Evans} {et~al.}(2001){Evans}, {Rawlings}, {Shirley}, \&
  {Mundy}}]{evans2001}
{Evans}, II, N.~J., {Rawlings}, J.~M.~C., {Shirley}, Y.~L., \& {Mundy}, L.~G.
  2001, \apj, 557, 193

\bibitem[{{Evans} {et~al.}(2003){Evans}, {Allen}, {Blake}, {Boogert}, {Bourke},
  {Harvey}, {Kessler}, {Koerner}, {Lee}, {Mundy}, {Myers}, {Padgett},
  {Pontoppidan}, {Sargent}, {Stapelfeldt}, {van Dishoeck}, {Young}, \&
  {Young}}]{evans2003}
{Evans}, II, N.~J., {Allen}, L.~E., {Blake}, G.~A., {et~al.} 2003, \pasp, 115,
  965

\bibitem[{{Fazio} {et~al.}(2004){Fazio}, {Hora}, {Allen}, {Ashby}, {Barmby},
  {Deutsch}, {Huang}, {Kleiner}, {Marengo}, {Megeath}, {Melnick}, {Pahre},
  {Patten}, {Polizotti}, {Smith}, {Taylor}, {Wang}, {Willner}, {Hoffmann},
  {Pipher}, {Forrest}, {McMurty}, {McCreight}, {McKelvey}, {McMurray}, {Koch},
  {Moseley}, {Arendt}, {Mentzell}, {Marx}, {Losch}, {Mayman}, {Eichhorn},
  {Krebs}, {Jhabvala}, {Gezari}, {Fixsen}, {Flores}, {Shakoorzadeh}, {Jungo},
  {Hakun}, {Workman}, {Karpati}, {Kichak}, {Whitley}, {Mann}, {Tollestrup},
  {Eisenhardt}, {Stern}, {Gorjian}, {Bhattacharya}, {Carey}, {Nelson},
  {Glaccum}, {Lacy}, {Lowrance}, {Laine}, {Reach}, {Stauffer}, {Surace},
  {Wilson}, {Wright}, {Hoffman}, {Domingo}, \& {Cohen}}]{fazio2004}
{Fazio}, G.~G., {Hora}, J.~L., {Allen}, L.~E., {et~al.} 2004, \apjs, 154, 10

\bibitem[{{Flaherty} {et~al.}(2007){Flaherty}, {Pipher}, {Megeath}, {Winston},
  {Gutermuth}, {Muzerolle}, {Allen}, \& {Fazio}}]{flaherty2007}
{Flaherty}, K.~M., {Pipher}, J.~L., {Megeath}, S.~T., {et~al.} 2007, \apj, 663,
  1069

\bibitem[{{Forbrich} {et~al.}(2010){Forbrich}, {Tappe}, {Robitaille}, {Muench},
  {Teixeira}, {Lada}, {Stolte}, \& {Lada}}]{forbrich2010}
{Forbrich}, J., {Tappe}, A., {Robitaille}, T., {et~al.} 2010, \apj, 716, 1453

\bibitem[{{Furlan} {et~al.}(2016){Furlan}, {Fischer}, {Ali}, {Stutz}, {Stanke},
  {Tobin}, {Megeath}, {Osorio}, {Hartmann}, {Calvet}, {Poteet}, {Booker},
  {Manoj}, {Watson}, \& {Allen}}]{furlan2016}
{Furlan}, E., {Fischer}, W.~J., {Ali}, B., {et~al.} 2016, \apjs, 224, 5

\bibitem[{{Gramajo} {et~al.}(2010){Gramajo}, {Whitney}, {G{\'o}mez}, \&
  {Robitaille}}]{gramajo2010}
{Gramajo}, L.~V., {Whitney}, B.~A., {G{\'o}mez}, M., \& {Robitaille}, T.~P.
  2010, \aj, 139, 2504

\bibitem[{{Greene} {et~al.}(1994){Greene}, {Wilking}, {Andre}, {Young}, \&
  {Lada}}]{greene1994}
{Greene}, T.~P., {Wilking}, B.~A., {Andre}, P., {Young}, E.~T., \& {Lada},
  C.~J. 1994, \apj, 434, 614

\bibitem[{{Haisch} {et~al.}(2006){Haisch}, {Barsony}, {Ressler}, \&
  {Greene}}]{haisch2006}
{Haisch}, Jr., K.~E., {Barsony}, M., {Ressler}, M.~E., \& {Greene}, T.~P. 2006,
  \aj, 132, 2675

\bibitem[{{Hartmann}(1998)}]{hartmann1998}
{Hartmann}, L. 1998, {Accretion Processes in Star Formation}

\bibitem[{{Harvey} {et~al.}(2007){Harvey}, {Mer{\'{\i}}n}, {Huard}, {Rebull},
  {Chapman}, {Evans}, \& {Myers}}]{harvey2007}
{Harvey}, P., {Mer{\'{\i}}n}, B., {Huard}, T.~L., {et~al.} 2007, \apj, 663,
  1149

\bibitem[{{Harvey} {et~al.}(2013){Harvey}, {Fallscheer}, {Ginsburg}, {Terebey},
  {Andr{\'e}}, {Bourke}, {Di Francesco}, {K{\"o}nyves}, {Matthews}, \&
  {Peterson}}]{harvey2013}
{Harvey}, P.~M., {Fallscheer}, C., {Ginsburg}, A., {et~al.} 2013, \apj, 764,
  133

\bibitem[{{Hatchell} {et~al.}(2007){Hatchell}, {Fuller}, {Richer}, {Harries},
  \& {Ladd}}]{hatchell2007}
{Hatchell}, J., {Fuller}, G.~A., {Richer}, J.~S., {Harries}, T.~J., \& {Ladd},
  E.~F. 2007, \aap, 468, 1009

\bibitem[{{Herter} {et~al.}(2012){Herter}, {Adams}, {De Buizer}, {Gull},
  {Schoenwald}, {Henderson}, {Keller}, {Nikola}, {Stacey}, \&
  {Vacca}}]{herter2012}
{Herter}, T.~L., {Adams}, J.~D., {De Buizer}, J.~M., {et~al.} 2012, \apjl, 749,
  L18

\bibitem[{{Horn} \& {Becklin}(2001)}]{horn2001}
{Horn}, J.~M.~M., \& {Becklin}, E.~E. 2001, \pasp, 113, 997

\bibitem[{{Indebetouw} {et~al.}(2005){Indebetouw}, {Mathis}, {Babler}, {Meade},
  {Watson}, {Whitney}, {Wolff}, {Wolfire}, {Cohen}, {Bania}, {Benjamin},
  {Clemens}, {Dickey}, {Jackson}, {Kobulnicky}, {Marston}, {Mercer},
  {Stauffer}, {Stolovy}, \& {Churchwell}}]{indebetouw2005}
{Indebetouw}, R., {Mathis}, J.~S., {Babler}, B.~L., {et~al.} 2005, \apj, 619,
  931

\bibitem[{{Kim} {et~al.}(1994){Kim}, {Martin}, \& {Hendry}}]{kmh1994}
{Kim}, S.-H., {Martin}, P.~G., \& {Hendry}, P.~D. 1994, \apj, 422, 164

\bibitem[{{Kryukova} {et~al.}(2012){Kryukova}, {Megeath}, {Gutermuth},
  {Pipher}, {Allen}, {Allen}, {Myers}, \& {Muzerolle}}]{kryukova2012}
{Kryukova}, E., {Megeath}, S.~T., {Gutermuth}, R.~A., {et~al.} 2012, \aj, 144,
  31

\bibitem[{{Lada}(1987)}]{lada1987}
{Lada}, C.~J. 1987, in IAU Symposium, Vol. 115, Star Forming Regions, ed.
  M.~{Peimbert} \& J.~{Jugaku}, 1--17

\bibitem[{{Lazareff} {et~al.}(1990){Lazareff}, {Monin}, \&
  {Pudritz}}]{lazareff1990}
{Lazareff}, B., {Monin}, J.-L., \& {Pudritz}, R.~E. 1990, \apj, 358, 170

\bibitem[{{Mathis} {et~al.}(1983){Mathis}, {Mezger}, \& {Panagia}}]{mathis1983}
{Mathis}, J.~S., {Mezger}, P.~G., \& {Panagia}, N. 1983, \aap, 128, 212

\bibitem[{{Maury} {et~al.}(2011){Maury}, {Andr{\'e}}, {Men'shchikov},
  {K{\"o}nyves}, \& {Bontemps}}]{maury2011}
{Maury}, A.~J., {Andr{\'e}}, P., {Men'shchikov}, A., {K{\"o}nyves}, V., \&
  {Bontemps}, S. 2011, \aap, 535, A77

\bibitem[{{Megeath} {et~al.}(2012){Megeath}, {Gutermuth}, {Muzerolle},
  {Kryukova}, {Flaherty}, {Hora}, {Allen}, {Hartmann}, {Myers}, {Pipher},
  {Stauffer}, {Young}, \& {Fazio}}]{megeath2012}
{Megeath}, S.~T., {Gutermuth}, R., {Muzerolle}, J., {et~al.} 2012, \aj, 144,
  192

\bibitem[{{Mer{\'{\i}}n} {et~al.}(2008){Mer{\'{\i}}n}, {J{\o}rgensen},
  {Spezzi}, {Alcal{\'a}}, {Evans}, {Harvey}, {Prusti}, {Chapman}, {Huard}, {van
  Dishoeck}, \& {Comer{\'o}n}}]{merin2008}
{Mer{\'{\i}}n}, B., {J{\o}rgensen}, J., {Spezzi}, L., {et~al.} 2008, \apjs,
  177, 551

\bibitem[{{Ormel} {et~al.}(2011){Ormel}, {Min}, {Tielens}, {Dominik}, \&
  {Paszun}}]{ormel2011}
{Ormel}, C.~W., {Min}, M., {Tielens}, A.~G.~G.~M., {Dominik}, C., \& {Paszun},
  D. 2011, \aap, 532, A43

\bibitem[{{Ossenkopf} \& {Henning}(1994)}]{oh1994}
{Ossenkopf}, V., \& {Henning}, T. 1994, \aap, 291, 943

\bibitem[{{Pollack} {et~al.}(1994){Pollack}, {Hollenbach}, {Beckwith},
  {Simonelli}, {Roush}, \& {Fong}}]{pollack1994}
{Pollack}, J.~B., {Hollenbach}, D., {Beckwith}, S., {et~al.} 1994, \apj, 421,
  615

\bibitem[{{Poulton} {et~al.}(2008){Poulton}, {Robitaille}, {Greaves},
  {Bonnell}, {Williams}, \& {Heyer}}]{poulton2008}
{Poulton}, C.~J., {Robitaille}, T.~P., {Greaves}, J.~S., {et~al.} 2008, \mnras,
  384, 1249

\bibitem[{{Pringle}(1981)}]{pringle1981}
{Pringle}, J.~E. 1981, \araa, 19, 137

\bibitem[{{Rebull} {et~al.}(2010){Rebull}, {Padgett}, {McCabe}, {Hillenbrand},
  {Stapelfeldt}, {Noriega-Crespo}, {Carey}, {Brooke}, {Huard}, {Terebey},
  {Audard}, {Monin}, {Fukagawa}, {G{\"u}del}, {Knapp}, {Menard}, {Allen},
  {Angione}, {Baldovin-Saavedra}, {Bouvier}, {Briggs}, {Dougados}, {Evans},
  {Flagey}, {Guieu}, {Grosso}, {Glauser}, {Harvey}, {Hines}, {Latter},
  {Skinner}, {Strom}, {Tromp}, \& {Wolf}}]{rebull2010}
{Rebull}, L.~M., {Padgett}, D.~L., {McCabe}, C.-E., {et~al.} 2010, \apjs, 186,
  259

\bibitem[{{Rieke} {et~al.}(2004){Rieke}, {Young}, {Engelbracht}, {Kelly},
  {Low}, {Haller}, {Beeman}, {Gordon}, {Stansberry}, {Misselt}, {Cadien},
  {Morrison}, {Rivlis}, {Latter}, {Noriega-Crespo}, {Padgett}, {Stapelfeldt},
  {Hines}, {Egami}, {Muzerolle}, {Alonso-Herrero}, {Blaylock}, {Dole}, {Hinz},
  {Le Floc'h}, {Papovich}, {P{\'e}rez-Gonz{\'a}lez}, {Smith}, {Su}, {Bennett},
  {Frayer}, {Henderson}, {Lu}, {Masci}, {Pesenson}, {Rebull}, {Rho}, {Keene},
  {Stolovy}, {Wachter}, {Wheaton}, {Werner}, \& {Richards}}]{rieke2004}
{Rieke}, G.~H., {Young}, E.~T., {Engelbracht}, C.~W., {et~al.} 2004, \apjs,
  154, 25

\bibitem[{{Sadavoy} {et~al.}(2014){Sadavoy}, {Di Francesco}, {Andr{\'e}},
  {Pezzuto}, {Bernard}, {Maury}, {Men'shchikov}, {Motte},
  {Nguy{\tilde}{\^e}n-Lu'o'ng}, {Schneider}, {Arzoumanian}, {Benedettini},
  {Bontemps}, {Elia}, {Hennemann}, {Hill}, {K{\"o}nyves}, {Louvet}, {Peretto},
  {Roy}, \& {White}}]{sadavoy2014}
{Sadavoy}, S.~I., {Di Francesco}, J., {Andr{\'e}}, P., {et~al.} 2014, \apjl,
  787, L18

\bibitem[{{Samal} {et~al.}(2012){Samal}, {Pandey}, {Ojha}, {Chauhan}, {Jose},
  \& {Pandey}}]{samal2012}
{Samal}, M.~R., {Pandey}, A.~K., {Ojha}, D.~K., {et~al.} 2012, \apj, 755, 20

\bibitem[{{Seale} \& {Looney}(2008)}]{seale2008}
{Seale}, J.~P., \& {Looney}, L.~W. 2008, \apj, 675, 427

\bibitem[{{Shakura} \& {Sunyaev}(1973)}]{shakura1973}
{Shakura}, N.~I., \& {Sunyaev}, R.~A. 1973, \aap, 24, 337

\bibitem[{{Shirley} {et~al.}(2011){Shirley}, {Huard}, {Pontoppidan}, {Wilner},
  {Stutz}, {Bieging}, \& {Evans}}]{shirley2011}
{Shirley}, Y.~L., {Huard}, T.~L., {Pontoppidan}, K.~M., {et~al.} 2011, \apj,
  728, 143

\bibitem[{{Shirley} {et~al.}(2005){Shirley}, {Nordhaus}, {Grcevich}, {Evans},
  {Rawlings}, \& {Tatematsu}}]{shirley2005}
{Shirley}, Y.~L., {Nordhaus}, M.~K., {Grcevich}, J.~M., {et~al.} 2005, \apj,
  632, 982

\bibitem[{{Shu}(1977)}]{shu1977}
{Shu}, F.~H. 1977, \apj, 214, 488

\bibitem[{{Simon} {et~al.}(2007){Simon}, {Bolatto}, {Whitney}, {Robitaille},
  {Shah}, {Makovoz}, {Stanimirovi{\'c}}, {Barb{\'a}}, \& {Rubio}}]{simon2007}
{Simon}, J.~D., {Bolatto}, A.~D., {Whitney}, B.~A., {et~al.} 2007, \apj, 669,
  327

\bibitem[{{Stutz} {et~al.}(2013){Stutz}, {Tobin}, {Stanke}, {Megeath},
  {Fischer}, {Robitaille}, {Henning}, {Ali}, {di Francesco}, {Furlan},
  {Hartmann}, {Osorio}, {Wilson}, {Allen}, {Krause}, \& {Manoj}}]{stutz2013}
{Stutz}, A.~M., {Tobin}, J.~J., {Stanke}, T., {et~al.} 2013, \apj, 767, 36

\bibitem[{{Suutarinen} {et~al.}(2013){Suutarinen}, {Haikala}, {Harju},
  {Juvela}, {Andr{\'e}}, {Kirk}, {K{\"o}nyves}, \& {White}}]{suut2013}
{Suutarinen}, A., {Haikala}, L.~K., {Harju}, J., {et~al.} 2013, \aap, 555, A140

\bibitem[{{Terebey} {et~al.}(1984){Terebey}, {Shu}, \& {Cassen}}]{terebey1984}
{Terebey}, S., {Shu}, F.~H., \& {Cassen}, P. 1984, \apj, 286, 529

\bibitem[{{Terebey} {et~al.}(2009){Terebey}, {Fich}, {Noriega-Crespo},
  {Padgett}, {Fukagawa}, {Audard}, {Brooke}, {Carey}, {Evans}, {Guedel},
  {Hines}, {Huard}, {Knapp}, {McCabe}, {Menard}, {Monin}, \&
  {Rebull}}]{terebey2009}
{Terebey}, S., {Fich}, M., {Noriega-Crespo}, A., {et~al.} 2009, \apj, 696, 1918

\bibitem[{{Tobin} {et~al.}(2007){Tobin}, {Looney}, {Mundy}, {Kwon}, \&
  {Hamidouche}}]{tobin2007}
{Tobin}, J.~J., {Looney}, L.~W., {Mundy}, L.~G., {Kwon}, W., \& {Hamidouche},
  M. 2007, \apj, 659, 1404

\bibitem[{{Ulrich}(1976)}]{ulrich1976}
{Ulrich}, R.~K. 1976, \apj, 210, 377

\bibitem[{{Whitney} \& {Hartmann}(1993)}]{whitney1993}
{Whitney}, B.~A., \& {Hartmann}, L. 1993, \apj, 402, 605

\bibitem[{{Whitney} {et~al.}(2004){Whitney}, {Indebetouw}, {Bjorkman}, \&
  {Wood}}]{whitney2004b}
{Whitney}, B.~A., {Indebetouw}, R., {Bjorkman}, J.~E., \& {Wood}, K. 2004,
  \apj, 617, 1177

\bibitem[{{Whitney} {et~al.}(2013){Whitney}, {Robitaille}, {Bjorkman}, {Dong},
  {Wolff}, {Wood}, \& {Honor}}]{whitney2013}
{Whitney}, B.~A., {Robitaille}, T.~P., {Bjorkman}, J.~E., {et~al.} 2013, \apjs,
  207, 30

\bibitem[{{Whitney} {et~al.}(2003{\natexlab{a}}){Whitney}, {Wood}, {Bjorkman},
  \& {Cohen}}]{whitney2003b}
{Whitney}, B.~A., {Wood}, K., {Bjorkman}, J.~E., \& {Cohen}, M.
  2003{\natexlab{a}}, \apj, 598, 1079

\bibitem[{{Whitney} {et~al.}(2003{\natexlab{b}}){Whitney}, {Wood}, {Bjorkman},
  \& {Wolff}}]{whitney2003a}
{Whitney}, B.~A., {Wood}, K., {Bjorkman}, J.~E., \& {Wolff}, M.~J.
  2003{\natexlab{b}}, \apj, 591, 1049

\bibitem[{{Whitney} {et~al.}(2008){Whitney}, {Sewilo}, {Indebetouw},
  {Robitaille}, {Meixner}, {Gordon}, {Meade}, {Babler}, {Harris}, {Hora},
  {Bracker}, {Povich}, {Churchwell}, {Engelbracht}, {For}, {Block}, {Misselt},
  {Vijh}, {Leitherer}, {Kawamura}, {Blum}, {Cohen}, {Fukui}, {Mizuno},
  {Mizuno}, {Srinivasan}, {Tielens}, {Volk}, {Bernard}, {Boulanger}, {Frogel},
  {Gallagher}, {Gorjian}, {Kelly}, {Latter}, {Madden}, {Kemper}, {Mould},
  {Nota}, {Oey}, {Olsen}, {Onishi}, {Paladini}, {Panagia}, {Perez-Gonzalez},
  {Reach}, {Shibai}, {Sato}, {Smith}, {Staveley-Smith}, {Ueta}, {Van Dyk},
  {Werner}, {Wolff}, \& {Zaritsky}}]{whitney2008}
{Whitney}, B.~A., {Sewilo}, M., {Indebetouw}, R., {et~al.} 2008, \aj, 136, 18

\bibitem[{{Wood} {et~al.}(1996){Wood}, {Bjorkman}, {Whitney}, \&
  {Code}}]{wood1996}
{Wood}, K., {Bjorkman}, J.~E., {Whitney}, B.~A., \& {Code}, A.~D. 1996, \apj,
  461, 828

\bibitem[{{Wood} {et~al.}(2002){Wood}, {Wolff}, {Bjorkman}, \&
  {Whitney}}]{wood2002}
{Wood}, K., {Wolff}, M.~J., {Bjorkman}, J.~E., \& {Whitney}, B. 2002, \apj,
  564, 887

\bibitem[{{Yorke} \& {Bodenheimer}(1999)}]{yorke1999}
{Yorke}, H.~W., \& {Bodenheimer}, P. 1999, \apj, 525, 330

\bibitem[{{Young} {et~al.}(2012){Young}, {Becklin}, {Marcum}, {Roellig}, {De
  Buizer}, {Herter}, {G{\"u}sten}, {Dunham}, {Temi}, {Andersson}, {Backman},
  {Burgdorf}, {Caroff}, {Casey}, {Davidson}, {Erickson}, {Gehrz}, {Harper},
  {Harvey}, {Helton}, {Horner}, {Howard}, {Klein}, {Krabbe}, {McLean}, {Meyer},
  {Miles}, {Morris}, {Reach}, {Rho}, {Richter}, {Roeser}, {Sandell}, {Sankrit},
  {Savage}, {Smith}, {Shuping}, {Vacca}, {Vaillancourt}, {Wolf}, \&
  {Zinnecker}}]{young2012}
{Young}, E.~T., {Becklin}, E.~E., {Marcum}, P.~M., {et~al.} 2012, \apjl, 749,
  L17

\bibitem[{{Young} {et~al.}(2005){Young}, {Harvey}, {Brooke}, {Chapman},
  {Kauffmann}, {Bertoldi}, {Lai}, {Alcal{\'a}}, {Bourke}, {Spiesman}, {Allen},
  {Blake}, {Evans}, {Koerner}, {Mundy}, {Myers}, {Padgett}, {Salinas},
  {Sargent}, {Stapelfeldt}, {Teuben}, {van Dishoeck}, \& {Wahhaj}}]{Kyoung2005}
{Young}, K.~E., {Harvey}, P.~M., {Brooke}, T.~Y., {et~al.} 2005, \apj, 628, 283

\bibitem[{{Zhang} {et~al.}(2013){Zhang}, {Tan}, \& {McKee}}]{zhang2013}
{Zhang}, Y., {Tan}, J.~C., \& {McKee}, C.~F. 2013, \apj, 766, 86

\end{thebibliography}

\end{document}